\documentclass[10pt]{iopart}
\usepackage{iopams,mathrsfs,graphicx,hyperref,color,bm,lipsum,xspace}
\expandafter\let\csname equation*\endcsname\relax
\expandafter\let\csname endequation*\endcsname\relax
\newcommand{\gae}{\lower 2pt \hbox{$\, \buildrel {\scriptstyle >}\over {\scriptstyle \sim}\,$}}
\newcommand{\lae}{\lower 2pt \hbox{$\, \buildrel {\scriptstyle <}\over {\scriptstyle \sim}\,$}}
\usepackage{amsmath,physics,soul}

\usepackage{etoolbox}
\makeatletter
\newrobustcmd{\fixappendix}{%
  \patchcmd{\l@section}{1.5em}{7em}{}{}%
  \patchcmd{\l@subsection}{2.3em}{7em}{}{}%
}
\makeatother

\newcommand{\ee}{\mathrm{e}}
\newcommand{\tA}{{Type A}\xspace}
\newcommand{\tB}{{Type B}\xspace}
\newcommand{\tAa}{\textit{Type A}\xspace}
\newcommand{\tBb}{\textit{Type B}\xspace}
\newcommand{\ii}{\mathrm{i}}

\begin{document}
\title[Dynamics of lattice random walk within heterogeneous media]{Dynamics of lattice random walk within regions composed of different media and interfaces}

\author{Debraj Das}
\address{Department of Engineering Mathematics, University of
Bristol, Bristol BS8 1TW, United Kingdom}
\ead{debrajdasphys@gmail.com, debraj.das@bristol.ac.uk,}
\author{Luca Giuggioli}
\address{Bristol Centre for Complexity Sciences and Department of Engineering Mathematics, University of
Bristol, Bristol BS8 1TW, United Kingdom}
\ead{Luca.Giuggioli@bristol.ac.uk}

\vspace{10pt}
\begin{indented}
\item[]\today
\end{indented}

\begin{abstract}
We study the lattice random walk dynamics in a heterogeneous space of two media separated by an interface and having different diffusivity and bias. Depending on the position of the interface, there exist two exclusive ways to model the dynamics: (1) \tAa dynamics whereby the interface is placed between two lattice points, and (2) \tBb dynamics whereby the interface is placed on a lattice point. For both types, we obtain exact results for the one-dimensional generating function of the Green's function or propagator for the composite system in unbounded domain as well as domains confined with reflecting, absorbing, and mixed boundaries. For the case with reflecting confinement in the absence of bias, the steady-state probability shows a step-like behavior for the \tAa dynamics, while it is uniform for the \tBb dynamics. We also derive explicit expressions for the first-passage probability and the mean first-passage time, and compare the hitting time dependence to a single target. Finally, considering the continuous-space continuous-time limit of the propagator, we obtain the boundary conditions at the interface. At the interface, while the flux is the same, the probability density is discontinuous for \tAa and is continuous for \tBb. For the latter we derive a generalized version of the so-called leather boundary condition in the appropriate limit.
\end{abstract}

\maketitle

\section{Introduction}

Diffusion in heterogeneous space is a ubiquitous transport phenomenon observed in diverse natural and engineering processes from heat conduction in solids~\cite{carslaw_conduction_1986} and muscles~\cite{gilbert_analysis_1988}, 
drug-delivery in tissues~\cite{pontrelli_mass_2007,todo_mathematical_2013} and invasion of cancerous cells~\cite{mantzavinos_fokas_2016},
to synthesis of materials~\cite{frenkel_theorie_1924,basuki_decoupling_2014,faupel_diffusion_2003}, electrical performance in electrodes~\cite{diard_one_2005,freger_diffusion_2005} and semiconductors~\cite{aguirre_heat_2000,graff_mechanisms_2004}, and
molecular exchanges in biological systems~\cite{pontrelli_mass_2007,todo_mathematical_2013}.
Given such a wide range of applications it is expected that many different methodologies have been used over the years to understand what role heterogeneities have on the random movement of particles.
The continuous-space continuous-time (CSCT) representation has had an important role in this respect. 
Techniques such as finite difference methods~\cite{hickson_finite_2011}, spectral decomposition~\cite{hickson_critical_2009,hickson_critical_2009-1,moutal_diffusion_2019}, 
orthogonal expansion of the Laplacian eigenmodes~\cite{carr_semi-analytical_2016}, 
and stochastic Monte-Carlo simulations~\cite{lejay_simulating_2012,lejay_monte_2018} have been instrumental to gain important insights.
In the context of first-passage problems, heterogeneous diffusion in composite~\cite{redner_guide_2001} and spherically symmetric~\cite{vaccario_first-passage_2015,godec_optimization_2015,godec_first_2016} media has been studied extensively.
Recently, a new fundamental equation that generalizes the diffusion and the Smoluchowski equation has been derived~\cite{kay_diffusion_2022}.
It allows to describe the dynamics of a Brownian particle in presence of an arbitrary finite number of infinitesimally thin permeable interfaces.

For heterogeneous diffusion, the spatial heterogeneity present at the interface between regions with unequal diffusion constant is naturally accounted for by imposing boundary conditions to the equation governing the dynamics of the occupation probability.
But these boundary conditions are often chosen without a derivation of the underlying microscopic dynamics at the interface.
Arguably, in the literature, there are very few studies that attempt to derive through first principles these boundary conditions. 
In Refs.~\cite{korabel_paradoxes_2010,korabel_boundary_2011}, the infiltration of diffusing particles from one medium to another in presence of anomalous diffusion has been studied. 
Starting from a continuous-time random walk model, it has been shown that an averaged net drift opposite to the imposed flow is possible and leads to a boundary condition that shows that the probability density is discontinuous at the interface. 
Another approach~\cite{kosztolowicz_membrane_2001} shows that the flow across a membrane is proportional to the difference of the concentrations at the opposite sides of the membrane, which is nothing but the so-called leather boundary condition~\cite{powles_exact_1992}. 

Aside from CSCT models, there have been spatially discrete analyses to study the dynamics in presence of permeable barriers using periodically placed heterogeneities~\cite{powles_exact_1992,kenkre_molecular_2008}. 
Subdiffusive dynamics has been considered in presence of a thin membrane~\cite{kosztolowicz_random_2015,kosztolowicz_subdiffusion_2017}, where,
starting from a discrete-space discrete-time (DSDT) random walk model, the Green's function or propagator and a nontrivial boundary condition at the membrane involving the Riemann-Liouville fractional time derivative have been found.
In these studies, the waiting time and jump probabilities at two adjacent lattice sites are modified, while the dynamics at all other sites remains that of a symmetric random walk. The boundary conditions are then obtained by taking the continuous space-time limit of the DSDT propagator.
In Ref.~\cite{kosztolowicz_subdiffusion_2017}, the spatial heterogeneity is introduced with different subdiffusion constants that are defined in the continuous limit through different waiting time distributions on either regions of the adjacent sites.
While these discrete space models~\cite{powles_exact_1992,kenkre_molecular_2008,kosztolowicz_random_2015,kosztolowicz_subdiffusion_2017} are of practical utility, none of them derives the leather boundary condition.

Here we derive a first principle procedure to represent the DSDT dynamics of a biased random walker in a heterogeneous space of two media separated by an interface and having different diffusivities. 
We extend the lattice random walk formalism developed in Refs.~\cite{giuggioli_exact_2020,sarvaharman_closed-form_2020,das_discrete_2022} to heterogeneous space  
and show that depending on the position of the interface on a lattice, there exist two exclusive ways to model it: (1) {\tAa} interface placed between two lattice points belonging to different media, and (2) {\tBb} interface placed exactly on a lattice point shared by both media.  
Very recently, an analytic framework based on the DSDT random walk approach has been developed to study diffusion with inert heterogeneities~\cite{sarvaharman_particle-environment_2022}.
It involves the computation of determinants and appears convenient in dealing with a finite number of heterogeneities. 
However, a DSDT procedure, when an entire medium or a large region is different, that bypasses the computation of large determinants does not exist.
Extending the mathematical approach in Ref.~\cite{kosztolowicz_subdiffusion_2017} to heterogeneous space in presence of bias, we obtain exact analytical solutions for the unbounded propagator for both the \tA and \tB interfaces. 
We further obtain the propagator in domains confined by reflecting, absorbing, and mixed boundaries.

The paper is organized as follows. 
In Sec.~\ref{sec:the-model}\,, we introduce the DSDT random-walk dynamics in an unbounded heterogeneous space of two different media, and present the corresponding Master equations. We solve the Master equations and obtain the exact propagator for the \tA and \tB dynamics in Sec.~\ref{sec:tA}\, and Sec.~\ref{sec:tB}\,, respectively. 
In Sec.~\ref{sec:confined}\, we discuss the propagators in confined domains. Sec.~\ref{subsec:fp} deals with the first-passage probability in a reflecting domain and its mean to a single target for both types of interfaces. 
In Sec.~\ref{sec:conti}\,, we obtain the unbounded propagators in the continuous space-time limit and the corresponding boundary conditions at the interface. We derive a generalized version of the so-called leather boundary condition in Sec.~\ref{sec:leather}\, and draw our conclusions in Sec.~\ref{sec:conclu}.


\section{The Model}
\label{sec:the-model}
The DSDT dynamics of a one dimensional (1d) biased lattice random walker (BLRW) is conveniently described by two parameters~\cite{sarvaharman_closed-form_2020}: 
(i) diffusivity $q$ with $0 \leq q\leq 1$, and (ii) bias $g$ with $-1\leq g\leq 1$. The walker from site $n$ at time $t$ in the next timestep either moves to site $(n - 1)$
with probability $q(1 + g)/2$ or to site $(n + 1)$ with probability $q(1 - g)/2$ or does not move at all with probability $(1 - q)$. Hence, the BLRW dynamics follows the Master equation $P(n,t+1)=(1-q)P(n,t)+ [q(1 - g)/2] P(n-1,t) + [q(1 + g)/2] P(n+1,t)$ with $P(n,t)$ being the probability that the walker is at site $n$ at time $t$. The dynamics is biased towards negative $x$-axis for $g > 0$ and towards positive $x$-axis for $g<0$, while the magnitude $|g|$ controls the strength of
the bias.
We are interested in studying the BLRW dynamics in heterogeneous space of two different media (with different diffusivities and biases) separated by an interface.  We denote the diffusivity and the bias parameter in the $\mu$-th medium by $q_\mu ~(0 \leq q_\mu\leq 1)$ and $g_\mu ~(-1\leq g_\mu\leq 1)$, respectively. 

\subsection{\tA interface}
\label{sec:the-model-tA}

\begin{figure}[!htbp]
\centering
\includegraphics[scale=1.05]{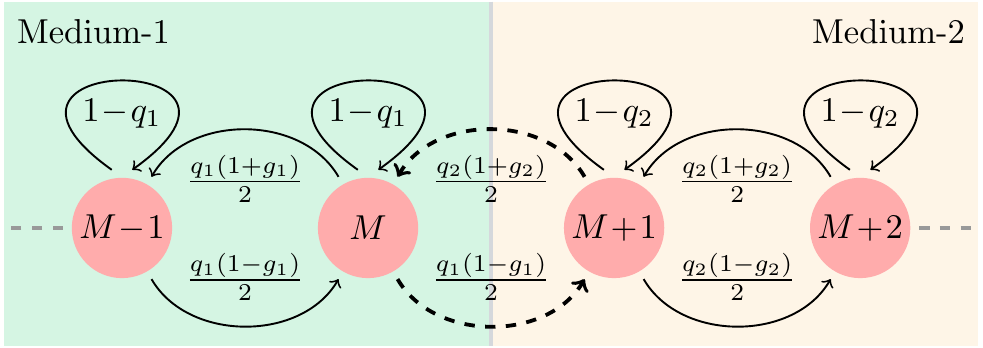}
\caption{Schematic diagram of the transition probabilities for the BLRW dynamics in heterogeneous space with two media separated by a \tA interface between sites $M$ and $(M+1)$.}
\label{fig:tA-dynamics}
\end{figure}

\noindent Figure~\ref{fig:tA-dynamics}\, schematically shows the transition probabilities for a 1d BLRW in an infinite heterogeneous space with two media separated by a \tA interface placed between sites $M$ and ($M+1$) such that sites $n \leq M$ and $n \geq (M+1)$ belong to the first and the second medium, respectively. Note that the inter-media transition probabilities (shown as dashed arrows in Fig.~\ref{fig:tA-dynamics}) are fixed by the normalization of the outgoing probabilities from sites $M$ and $(M+1)$.
Denoting the probability to find the walker at site $n$ in medium-$\mu$ at time $t$ by $P_{\mu}(n,t)$, one may write the Master equation for the \tA dynamics as
\begin{align}
 P_1 (n,t+1) &= (1-q_1) P_1(n,t) + \frac{q_1}{2} (1-g_1) P_1(n-1,t) \nonumber \\ 
& + \frac{q_1}{2} (1+g_1) P_1(n+1,t)   \, , \quad \quad \quad n < M  \, , \label{eq:master_eq_1} \\
 P_1 (M,t+1) &= (1-q_1) P_1(M,t) + \frac{q_1}{2} (1-g_1) P_1(M-1,t) \nonumber \\  
& + \frac{q_2}{2} (1+g_2) P_2(M+1,t)    \, , \label{eq:master_eq_2} \\
 P_2 (M+1,t+1) &= (1-q_2) P_2(M+1,t) + \frac{q_1}{2} (1-g_1) P_1(M,t) \nonumber \\  
& + \frac{q_2}{2} (1+g_2) P_2(M+2,t)  \, , \label{eq:master_eq_3} \\
 P_2 (n,t+1) &= (1-q_2) P_2(n,t) + \frac{q_2}{2} (1-g_2) P_2(n-1,t) \nonumber \\  
& + \frac{q_2}{2} (1+g_2) P_2(n+1,t)   \, , \quad \quad \quad n > (M+1) \, . \label{eq:master_eq_4}
\end{align}
The dynamics around the interface is governed by the two explicitly coupled Eqs.~\eqref{eq:master_eq_2} and~\eqref{eq:master_eq_3} that clearly connect the probabilities $P_1$ and $P_2$ at sites $M$ and $(M+1)$. The incoming probability to site $M$ in medium-1 from site $(M+1)$ in medium-2 is completely given in terms of the medium-2 parameters $q_2$ and $g_2$, and vice versa.

\subsection{\tB interface}
\label{sec:the-model-tB}

\begin{figure}[!htbp]
\centering
\includegraphics[scale=1.05]{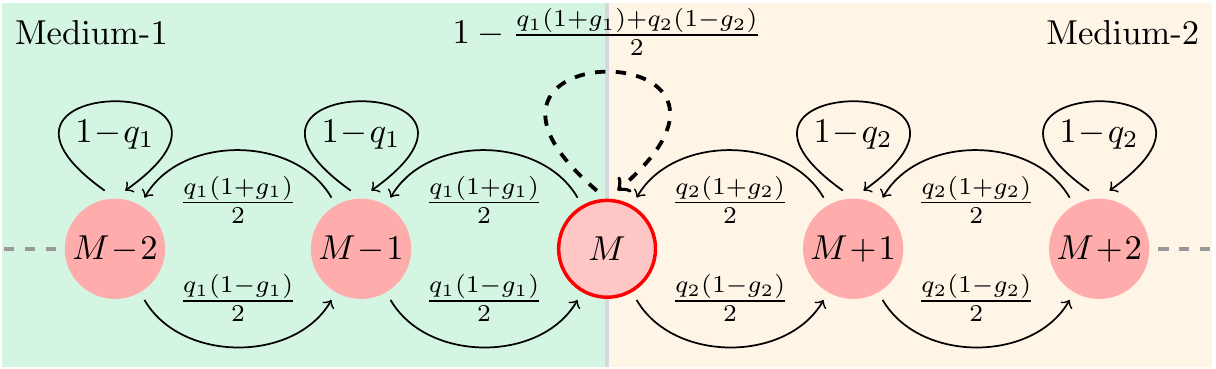}
\caption{Schematic diagram of the transition probabilities for the BLRW dynamics in heterogeneous space with two media separated by a \tB interface at the shared site $M$.}
\label{fig:tB-dynamics}
\end{figure}

\noindent Figure~\ref{fig:tB-dynamics}\, schematically shows the transition probabilities for a 1d BLRW in an infinite heterogeneous space with two media separated by a \tB interface placed at the shared site $M$. In this case, sites $n<M$ and $n>M$  belong to medium-1 and 2, respectively, while the interface at the shared site $M$ is associated with a hypothetical third medium-$c$. The probability of staying at the interface (shown as a dashed arrow in Fig.~\ref{fig:tB-dynamics}) is fixed by the normalization of the outgoing probabilities from it, and is subject to the constraint $q_c \equiv (q_1+q_2 +q_1g_1 - q_2g_2 )/2 \leq 1$, which is automatically satisfied in the absence of bias (as $g_1=g_2=0$). 
For the BLRW dynamics with the \tB interface, the Master equation is given by
\begin{align}
 P_1 (n,t+1) &= (1-q_1) P_1(n,t) + \frac{q_1}{2} (1-g_1)  P_1(n-1,t)  \nonumber \\ 
&  + \frac{q_1}{2} (1+g_1) P_1(n+1,t)   \, , \quad \quad \quad n < (M-1)  \, , \label{eq:master_eq_1_V} \\
 P_1 (M-1,t+1) &= (1-q_1) P_1(M-1,t) + \frac{q_1}{2} (1-g_1)  P_1(M-2,t)  \nonumber \\ 
&  + \frac{q_1}{2} (1+g_1) P_c(M,t)   \, , \label{eq:master_eq_2_V} \\
 P_{c} (M,t+1) &= \left[ 1 - \frac{q_1}{2} (1+g_1) - \frac{q_2}{2} (1-g_2) \right] P_c (M,t) + \frac{q_1}{2} (1-g_1) P_1(M-1,t) \nonumber \\
&  + \frac{q_2}{2} (1+g_2) P_2(M+1,t)    \, , \label{eq:master_eq_3_V}\\
 P_2 (M+1,t+1) &= (1-q_2) P_2(M+1,t) + \frac{q_2}{2} (1-g_2) P_c(M,t)  \nonumber \\ 
&  + \frac{q_2}{2} (1+g_2) P_2(M+2,t) \, , \label{eq:master_eq_4_V} \\
 P_2 (n,t+1) &= (1-q_2) P_2(n,t) + \frac{q_2}{2} (1-g_2) P_2(n-1,t)  \nonumber \\ 
& + \frac{q_2}{2} (1+g_2) P_2(n+1,t) \, , \quad \quad \quad n > (M+1) \, . \label{eq:master_eq_5_V} 
\end{align}
Unlike the case for \tA, the dynamics around the \tB interface yields three explicitly coupled Eqs.~\eqref{eq:master_eq_2_V}, \eqref{eq:master_eq_3_V}, and \eqref{eq:master_eq_4_V} that depict the mixing of the probabilities $P_1, P_c$, and $P_2$ at sites $(M-1),$ $M,$ and $(M+1)$ with $P_c$ representing the probability at site $M$. 
With the sharing of one site  between medium-1 and 2, the incoming probability to site $M$ and the outgoing probability from site $M$ are completely described the parameters $q_1$, $g_1$, $q_2$, and $g_2$. 

We emphasize that in a heterogeneous discrete space once the interface position, whether between two sites or on a site, is given, the transition probabilities between sites may not be chosen arbitrarily, rather they are fixed by the normalization of the outgoing probabilities. This very fact clearly suggests that depending on the position of the interface between two different media in heterogeneous space there exist two different ways to model the DSDT dynamics, namely, the \tA and the \tB dynamics. In both cases, the interface vanishes when $q_1=q_2$ and $g_1=g_2$, and the dynamics reduce to the one of the BLRW in homogeneous space studied in Ref.~\cite{sarvaharman_closed-form_2020}.

\section{Propagator in unbounded domain}
\subsection{\tA dynamics}
\label{sec:tA}

To solve the \tA Master equation~\eqref{eq:master_eq_1}--\eqref{eq:master_eq_4}, we use the following formalism. We denote the probability to find the walker at site $n$ in medium-$\mu$ at time $t$ while starting from site $n_0$ in medium-$\nu$ by the propagator $P_{\mu,\nu}(n,t|n_0)$.
By definition $P_{\mu,\nu}(n,t|n_0)$ vanishes when $n \notin$ medium-$\mu$ and/or $n_0 \notin$ medium-$\nu$.
In the $z$-domain, the propagator generating function is defined by $S_{\mu,\nu} (n,z|n_0) \equiv \sum_{t=0}^{\infty} \, z^t P_{\mu,\nu} (n,t|n_0)$. 
For simplicity, let us first consider that the walker starts from a site $n_0 \leq M$, i.e., $n_0 \in$ medium-1. 
In the $z$-domain, the \tA Master equation~\eqref{eq:master_eq_1}--\eqref{eq:master_eq_4}  may then be written as Eqs.~\eqref{eq:master_eq_inZ_1}--\eqref{eq:master_eq_inZ_4} given in~\ref{app:tA-solutions}.
To solve these equations we use the Fourier transform of $S_{\mu,\nu} (n,z|n_0)$, i.e.,
${\mathcal G}_{\mu,\nu} (k,z|n_0) \equiv \sum_{n=-\infty}^{\infty}  \exp(- \ii k n) \, S_{\mu,\nu} (n,z|n_0),$
such that its inverse is given by
$S_{\mu,\nu} (n,z|n_0) \equiv \int_{-\pi}^{\pi} \dd{k}   \exp( \ii k n ) \, \mathcal{G}_{\mu,\nu}(k,z|n_0) / (2\pi)$.
For later convenience, we introduce a complex variable $u \equiv \exp(-\ii k )$ so that the Fourier and the inverse transforms may be written as
\begin{align}
G_{\mu,\nu} (u,z|n_0) &\equiv \sum_{n=-\infty}^{\infty}  u^n \, S_{\mu,\nu} (n,z|n_0) \, , \label{eq:Fourier_def_in_u}\\
S_{\mu,\nu} (n,z|n_0) &\equiv \frac{1}{2 \pi \ii} \oint_{C} {\mathrm d}u ~ \frac{G_{\mu,\nu} (u,z|n_0) }{u^{n+1}}  \, , \label{eq:Inverse_Fourier_def_in_u}
\end{align}
where $C$ is the counterclockwise unit circular contour centered at $u=0$ on the complex $u$-plane. 
With two media and $n_0 \in $ medium-1, Eq.~\eqref{eq:Fourier_def_in_u} reduces to
\begin{align}
G_{1,1} (u,z|n_0) &\equiv \sum_{n=-\infty}^{M} u^n \, S_{1,1} (n,z|n_0) \, ,  \label{eq:GF_inU_1_def} \\
G_{2,1} (u,z|n_0) &\equiv \sum_{n=M+1}^{\infty} u^n \, S_{2,1} (n,z|n_0) \, . \label{eq:GF_inU_2_def}
\end{align}

The solution of the Master equation in the $z$-domain (Eqs.~\eqref{eq:master_eq_inZ_1}--\eqref{eq:master_eq_inZ_4}) is obtained as (see~\ref{app:tA-solutions})
\begin{align}
S_{1,1} (n,z|n_0) &= \frac{1}{\sqrt{1-\beta^{+}_1 \beta^{-}_1}} \Bigg[ \frac{ f_1^{\frac{n-n_0+|n-n_0|}{2}} \xi_1^{|n-n_0|}  }{ 1-z+z q_1 }  \nonumber \\
&\hskip10pt -\frac{\beta^{-}_1 (\xi_1  -  \xi_2) f_1^{M-n_0} \, \xi_1^{2 M-n -n_0}  }{\big(1-z+z q_1\big) \big( 2- \beta^{-}_1 \xi_1 - \beta^{-}_1  \xi_2 \big)}  \Bigg]  \, , \label{eq:GF_inZ_1_sol} 
\end{align}
\begin{align}
S_{2,1} (n,z|n_0) &= \frac{2 \, q_1(1-g_1) \big(f_1 \xi_1 \big)^{M-n_0} \big(f_2 \xi_2 \big)^{n-M} }{q_2 (1-g_2) \big(1-z+z q_1\big) \big( 2- \beta^{-}_1 \xi_1 - \beta^{-}_1  \xi_2 \big)}  \, , \label{eq:GF_inZ_2_sol}
\end{align}
where we have
\begin{align}
& \beta^{\pm}_\mu(z) \equiv  \frac{z q_\mu (1\pm g_\mu)}{1-z+ z q_\mu} \, , \label{eq:betaPM_i_def} \\
& f_\mu \equiv \frac{1-g_\mu}{1+g_\mu} = \frac{\beta^{-}_\mu}{\beta^{+}_\mu}  \, , \label{eq:f_i_def} \\
& \xi_\mu(z)  \equiv  \frac{1-\sqrt{1-\beta^{+}_\mu(z) \beta^{-}_\mu(z)}}{\beta^{-}_\mu(z)} = f^{-\frac{1}{2}}_\mu \Bigg( \frac{1}{\sqrt{\beta^{+}_\mu \beta^{-}_\mu}} - \sqrt{ \frac{1}{{\beta^{+}_\mu \beta^{-}_\mu}}  - 1} \,\Bigg) \, .   \label{eq:xi_i_def}
\end{align}
For $n_0 \in$ medium-2, i.e., with $n_0 \geq (M+1) $, following the same procedure one gets the solutions $S_{1,2}(n,z|n_0)$ and $S_{2,2}(n,z|n_0)$ given in Eqs.~\eqref{eq:GF_inZ_1_sol_Reg2} and~\eqref{eq:GF_inZ_2_sol_Reg2}, respectively.
Combining Eqs.~\eqref{eq:GF_inZ_1_sol}--\eqref{eq:GF_inZ_2_sol} and Eqs.~\eqref{eq:GF_inZ_1_sol_Reg2}--\eqref{eq:GF_inZ_2_sol_Reg2}, one may write the general solution (valid for any initial condition) for the \tA propagator generating function as
\begin{align}
S(n,z|n_0) &= \Theta(M-n_0) \Big[ \Theta(M-n) S_{1,1}(n,z|n_0) + \Theta(n-M-1)  S_{2,1}(n,z|n_0) \Big] \nonumber \\
& + \Theta(n_0-M-1) \Big[ \Theta(M-n) S_{1,2}(n,z|n_0) + \Theta(n-M-1)  S_{2,2}(n,z|n_0) \Big]  \, , \label{eq:GF_inZ_2reg_gen_sol}
\end{align}
where the discrete Heaviside function $\Theta(n)$ is defined by
\begin{align}
\Theta(n) \equiv 
\begin{cases} 
      1 \, , & n \geq 0 \, ,\\
      0 \, , & n < 0  \, .
\end{cases} \label{eq:Heaviside-def}
\end{align}

In homogeneous space ($q_\mu=q$) with the same bias ($g_\mu= g$), one has $\beta_\mu^{\pm} = \beta^{\pm}$, $f_\mu = f$, $\xi_\mu = \xi$. 
In this case, the \tA generating function $S_{\mu,\nu}(n,z|n_0) ~\forall ~\mu, \nu \in \{1,2\}$ become equal, and the general solution~\eqref{eq:GF_inZ_2reg_gen_sol} may be written as
\begin{align}
S (n,z|n_0) &= \frac{ f^{\frac{n-n_0+|n-n_0|}{2}} \xi^{|n-n_0|} }{\sqrt{1-\beta^+ \beta^-} \, \big( 1-z+z q \big)  } \,  , \label{eq:BLRW_shree}
\end{align}
which is the same quantity, expressed in a different form, as reported in Eq.~(2) of Ref.~\cite{sarvaharman_closed-form_2020}. In the absence of a bias $(g=0)$, Eq.~\eqref{eq:BLRW_shree} further reduces to the lazy lattice walk propagator~\cite{giuggioli_exact_2020}
$S (n,z|n_0) = \xi^{|n-n_0|} / [\sqrt{1-\beta^2 } \, \big( 1-z+z q \big)]   $
with $ \beta = z q / (1-z+zq) $ and $\xi = 1/\beta - \sqrt{1/\beta^2-1}$.

\begin{figure}[!htbp]
\centering
\includegraphics[scale=0.935]{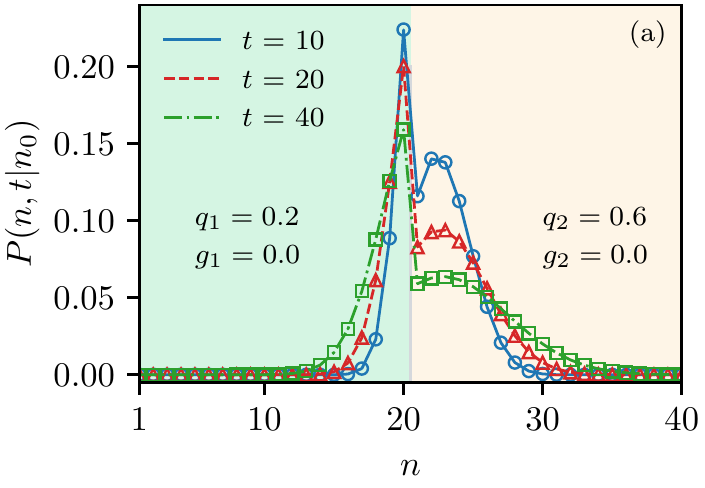}
\includegraphics[scale=0.935]{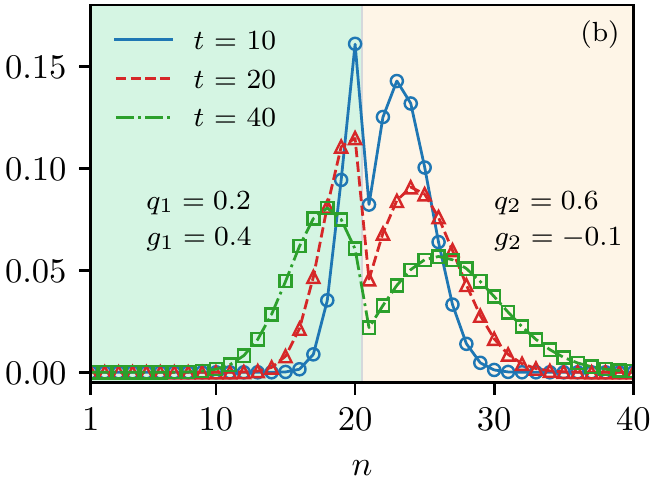}
\caption{Propagator $P(n,t|n_0)$ in the infinite domain with \tA interface between sites 20 and 21. The walker starts from site $n_0=22$ in medium-2 in the heterogeneous space with two different diffusivities $q_1=0.2$ and $q_2 = 0.6$. (a) Without bias. (b) With bias: $g_1=0.4,\, g_2=-0.1$. The lines are obtained by numerically inverting~\cite{abate_numerical_1992} Eq.~\eqref{eq:GF_inZ_2reg_gen_sol}, while the points are obtained from $5\times 10^6$ stochastic realizations. }
\label{fig:typeA-2M-Unbound-prop}
\end{figure}
Figure~\ref{fig:typeA-2M-Unbound-prop}\, shows the time-dependent propagator for the \tA dynamics when the walker starts in medium-2. Panel (a) depicts results in the absence of biases. As time increases the propagator shows a diffusive behavior with a discontinuity around the interface, which persists because of the unequal jump probabilities (that depend on the parameters from opposite media, see Fig.~\ref{fig:tA-dynamics}) between sites $M$ and $(M+1)$. Panel (b) depicts the propagator with opposite but unequal bias away from the interface (bias in medium-1 is four times larger). In this case, the smaller diffusivity in medium-1 makes the displacement of the peaks to the left smaller in magnitude than to the right.

\subsection{\tB dynamics}
\label{sec:tB}

Solving the \tB Master equation \eqref{eq:master_eq_1_V}--\eqref{eq:master_eq_5_V} using the same procedure (see~\ref{app:tB-solutions}), the generating functions with $n_0\leq (M-1)$, i.e., $n_0 \in$ medium-1 are obtained as
\begin{align}
&S_{1,1}(n, z|n_0) = \frac{1}{\sqrt{1-\beta^{+}_1  \beta^{-}_1}} \Bigg[ \frac{ f_1^{ \frac{ n - n_0 + |n-n_0|}{2} } \xi_1^{|n-n_0|} }{1-z + z q_1} \nonumber \\
&\hskip1.9cm  - f_1^{M-n_0} \, \xi_1^{2M-n-n_0} \bigg\{ \frac{1}{1-z+ z q_1} - \frac{1}{\Gamma} \, \sqrt{1-\beta^{+}_1  \beta^{-}_1} \bigg\} \Bigg]  \, , \label{eq:GF_inZ_S1_sol_V} \\
&S_{c,1}(M,z|n_0)  = \frac{1}{\Gamma} \, f_1^{M-n_0} \xi_1^{M-n_0} \, ,  \label{eq:GF_inZ_Sc_V} \\
&S_{2,1}(n, z|n_0) = \frac{1}{\Gamma} \, f_1^{M-n_0} \,   f_2^{n-M} \, \xi_1^{M-n_0}  \,  \xi_2^{n-M}  \, , \label{eq:GF_inZ_S2_sol_V}
\end{align}
where we have
\begin{align}
\Gamma \equiv 1-z + \frac{z q_1 }{2} \Big[ 1  + g_1 - \xi_1 (1- g_1) \Big]  + \frac{z q_2 }{2} \Big[1 - g_2 - f_2 \,  \xi_2 (1+g_2) \Big] \, . \label{eq:Gamma_def_V}
\end{align}
For other possible initial conditions, i.e., $n_0 =M$ and $n_0 \geq (M+1)$, the solutions $S_{\mu,\nu}(n, z|n_0); \, \mu \in \{1,c,2 \}, \, \nu \in\{c,2\},$ obtained following the same procedure, are reported in~\ref{app:tB-solutions}.
Combining all possible initial conditions, the general solution of the \tB propagator generating function may be written as
\begin{align}
S(n,z|n_0) &= \Theta(M-1-n_0) \Big[ \Theta(M-1-n) S_{1,1}(n,z|n_0) + \delta_{n,M}  S_{c,1}(n,z|n_0) \nonumber \\
& + \Theta(n-M-1)  S_{2,1}(n,z|n_0) \Big]  +\delta_{n_0,M}   \Big[ \Theta(M-1-n) S_{1,c}(n,z|n_0)  \nonumber \\ 
& + \delta_{n,M}  S_{c,c}(n,z|n_0) + \Theta(n-M-1)  S_{2,c}(n,z|n_0) \Big]  + \Theta(n_0-M-1)  \nonumber \\ 
& \times \Big[ \Theta(M-1-n) S_{1,2}(n,z|n_0) + \delta_{n,M}  S_{c,2}(n,z|n_0) \nonumber \\
& + \Theta(n-M-1)  S_{2,2}(n,z|n_0) \Big]   . \label{eq:GF_inZ_tB_gen_sol}
\end{align}
In homogeneous space ($q_\mu=q$) with the same bias ($g_\mu=g$), the \tB functions $S_{\mu,\nu}(n,z|n_0)$ $\forall ~\mu,\nu \in \{1,c,2 \}$ become equal, and the general solution~\eqref{eq:GF_inZ_tB_gen_sol} reduces to Eq.~\eqref{eq:BLRW_shree}.

\begin{figure}[!htbp]
\centering
\includegraphics[scale=0.935]{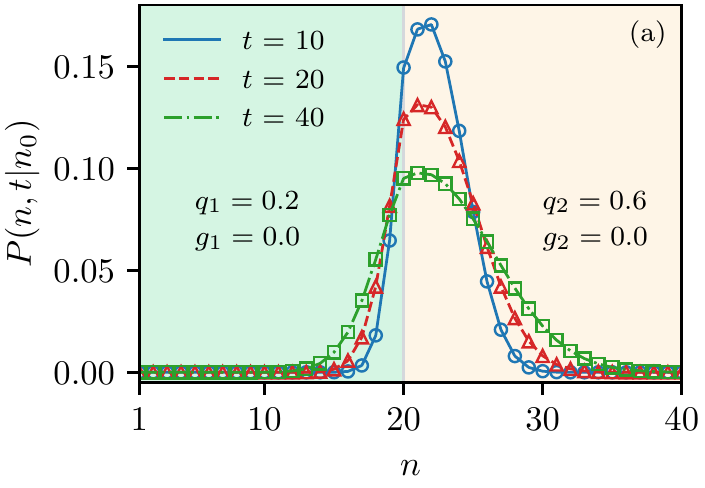}
\includegraphics[scale=0.935]{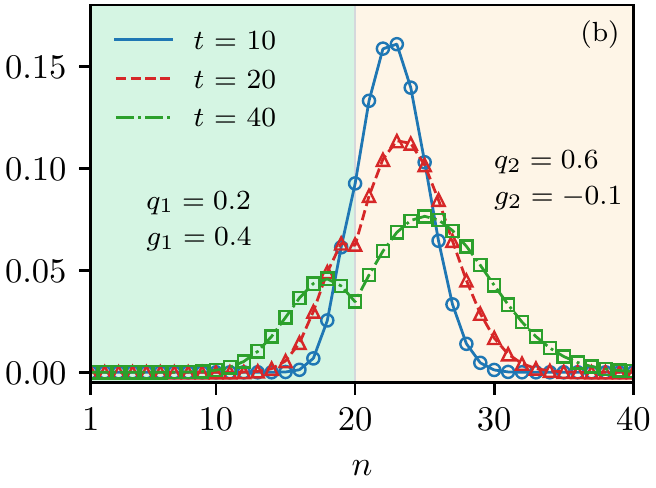}
\caption{Propagator $P(n,t|n_0)$ in the infinite domain with \tB interface at site 20: (a) without bias, and (b) with bias. Parameter values are the same as in Fig.~\eqref{fig:typeA-2M-Unbound-prop}. The lines are obtained by numerically inverting~\cite{abate_numerical_1992} Eq.~\eqref{eq:GF_inZ_tB_gen_sol}, while the points are obtained from $5\times 10^6$ stochastic realizations.}
\label{fig:typeB-2M-Unbound-prop}
\end{figure}

Figure~\ref{fig:typeB-2M-Unbound-prop}\, shows the time-dependent propagator for the \tB dynamics where the walker starts in medium-2. Panel (a) depicts results in the absence of biases. In this case, the diffusing propagator shape is smoother around the interface than the case of \tA (Fig.~\ref{fig:typeA-2M-Unbound-prop}(a)). 
This is because the interface is on the lattice site $M$ and the parameters from the two media only modifies the probability of staying at the interface. Panel (b) depicts the propagator with opposite but unequal bias away from the interface (bias in medium-1 is four times larger). 
Similar to the case for \tA we see that as $q_1<q_2$ the displacement of the peaks in medium-1 is smaller in magnitude than in medium-2.

\section{Propagator in confined domains}
\label{sec:confined}

We now consider the BLRW dynamics in a confined heterogeneous space bounded by barriers at sites $1$ and $N$ such that the interface (\tA or \tB) lies somewhere between the barriers. 
In solving the Master equation for confined lattice-walk dynamics in homogeneous space, the method of images is an intuitive and convenient technique.
It relies on translational invariance, which can be ensured also in presence of a global bias using an appropriate transformation~\cite{sarvaharman_closed-form_2020}.
However, for the dynamics in heterogeneous space translational invariance is lost, and it is not obvious how to use the method of images.       
We thus tackle the problem by employing the so-called Montroll's defect technique~\cite{montroll_random_1965,montroll_effect_1955,montroll_chapter_1979,kenkre_memory_2021}.
The technique exploits the linearity of Master equation in probability, seeks a solution assuming that a modification of the Master equation is given by a known term, and subsequently finds the actual solution of the problem with the defect by satisfying a self-consistent relation (Cramer's rule).
The procedure gives the exact solution of the Master equation, i.e., the defective propagator in terms of the defect-free propagator.

\begin{figure}[!htbp]
\centering
\includegraphics[scale=0.92]{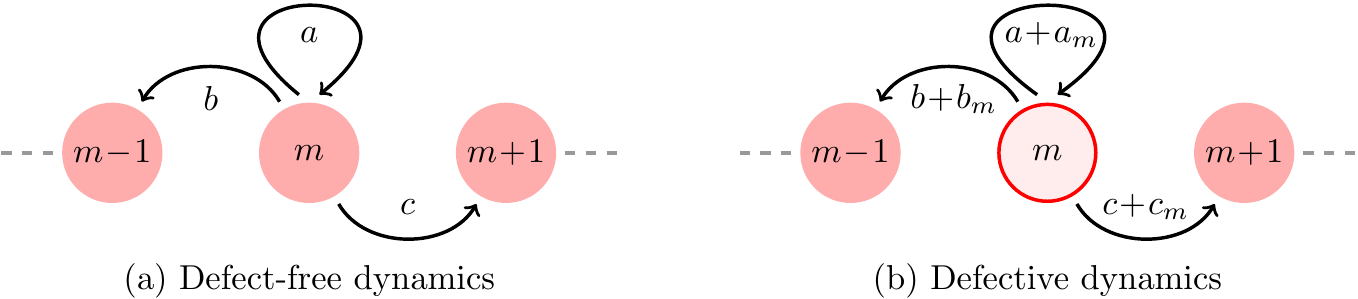}
\caption{Schematic diagram of the transition probabilities from a given site $n_1$. (a) The defect-free dynamics: The transition probabilities $a, \, b,\, c$. (b) Defective dynamics: The transition probabilities are modified with $a_m$, $b_m$, and $c_m$.  }
\label{fig:defect-1}
\end{figure}

To explain intuitively what the defect technique entails in discrete space we display a schematic representation in Fig.~\ref{fig:defect-1}.
For simplicity we consider a generic lattice walk dynamics with a single defective site and write the general form of the solution since it appears not to have been reported in the literature for discrete time.
Let us assume that $S(n,z|n_0)$ denotes the propagator generating function for the defect-free dynamics, in which the outgoing transition probabilities from a  site $m$ to sites $m,$ $(m-1),$ and $(m+1)$ are, respectively, given by $a$, $b$, and $c$, as shown in Fig.~\ref{fig:defect-1}(a). When these transition probabilities are modified, the site $m$ may be considered as a defective site, and correspondingly the dynamics become a defective one. In this defective dynamics, let us denote the modified transition probabilities from site $m$ to sites $(m-1),$ $m,$ and $(m+1)$ by  $(a+a_{m}),$ $(b+b_{m}),$ and $(c+c_{m})$, respectively, see Fig.~\ref{fig:defect-1}(b). Using the defect technique, one obtains an exact solution for the defective propagator generating function $P(n,z|n_0)$ in terms of the defect-free function $S(n,z|n_0)$ as
\begin{align}
P(n,z|n_0) &=  S(n,z|n_0) \nonumber \\
&+   \frac{z  \,  S(m,z|n_0) \big[  a_{m} S(n,z|m) + b_{m}  S(n,z|m-1) +  c_{m} S(n,z|m + 1)  \big]  }{1-z \big[  a_{m} S(m,z|m) + b_{m}  S(m,z|m-1) +  c_{m} S(m,z|m+1)  \big]} \, . \label{eq:Defect-1}
\end{align}
We note that in solving a problem using the defect technique, one does not require to use the normalization of probabilities, and hence it is also applicable when the defect-free jump probabilities are not conserved, i.e., the construct is valid for $(a+b+c)\leq 1$. 
By using Eq.~\eqref{eq:Defect-1} for one boundary at a time, the modification of the diffusive dynamics due to the boundaries at sites 1 and $N$ can be solved in terms of the unbounded dynamics.

\subsection{Reflecting boundaries}
\label{subsec:refl-prop}

\begin{figure}[t]
\centering
\includegraphics[scale=0.92]{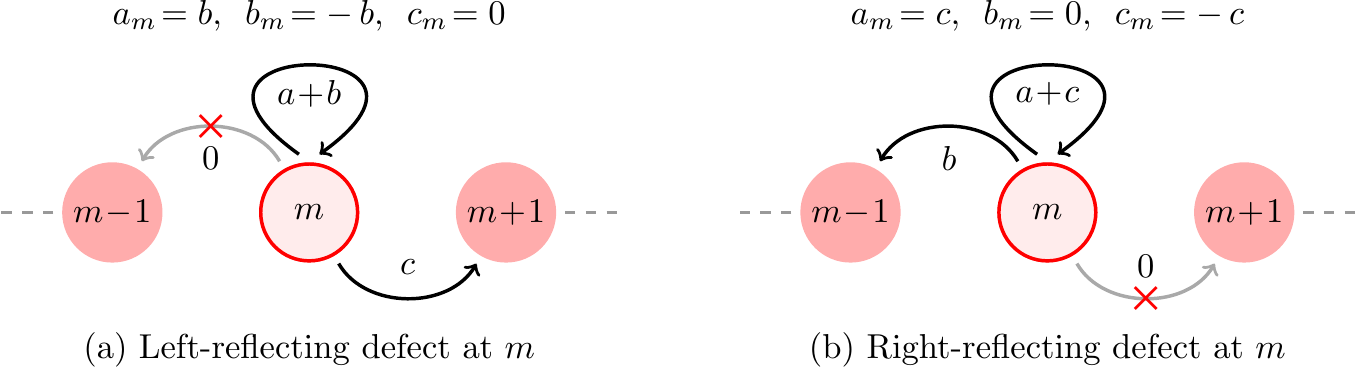}
\caption{Schematic diagram to show reflecting barriers as defects. (a) Left-reflecting defect: There exists a reflecting barrier at $m$ such that probability current from $m$ to $(m-1)$ vanishes. (b) Right-reflecting defect: There exists a reflecting barrier at $m$ such that probability current from $m$ to $(m+1)$ vanishes. }
\label{fig:defect-2}
\end{figure}

If the defective site $m$ mimics a reflecting barrier on the left so that the probability current from $m$ to $(m-1)$ vanishes (see Fig.~\ref{fig:defect-2}(a)), the dynamics becomes semi-bounded from the left, and the corresponding propagator is obtained from Eq.~\eqref{eq:Defect-1} by putting $a_{m}=b,$ $ b_{m}=-b,$ and $ c_{m}=0$. 
As there is no leakage of probability at a reflecting boundary, one has $(a_m+b_m+c_m)=0$.
Hence, the left-bounded propagator generating function $L^{\mathrm{(r)}}(n,z|n_0)$ corresponding to the generic defect-free dynamics is then given by (the superscript $r$ stands for reflecting boundary)
\begin{align}
L^{\mathrm{(r)}}(n,z|n_0) =  S(n,z|n_0) +   \frac{z b \,  S(m,z|n_0) \big[   S(n,z|m) -  S(n,z|m-1)   \big]  }{1-z b \big[  S(m,z|m) - S(m,z|m-1)   \big]}  ; \quad n,n_0 \geq m  . \label{eq:Defect-1-LB}
\end{align}
Similarly, when the defective site $m$ mimics a reflecting barrier on the right, the probability current from $m$ to $(m+1)$ vanishes (see Fig.~\ref{fig:defect-2}(b)). In this case, the right-bounded generating function $R^{\mathrm{(r)}}(n,z|n_0)$ obtained from Eq.~\eqref{eq:Defect-1} with $a_{m}=c,$ $ b_{m}=0,$ $ c_{m}=-c$ is given by
\begin{align}
R^{\mathrm{(r)}}(n,z|n_0) =  S(n,z|n_0) +   \frac{z c \,  S(m,z|n_0) \big[  S(n,z|m)  - S(n,z|m + 1)  \big]  }{1-z c \big[  S(m,z|m)  - S(m,z|m+1)  \big]} ; \quad n,n_0 \leq m  . \label{eq:Defect-1-RB}
\end{align}
In homogeneous space with a global bias, the semi-bounded reflecting propagator can also be obtained by using a transformation to symmetrize the dynamics and then by employing the method of images~\cite{sarvaharman_closed-form_2020}. 
Equations~\eqref{eq:Defect-1}--\eqref{eq:Defect-1-RB} bypass these cumbersome steps and also act as a working formula that can be readily used once the biased unbounded propagator is known. 

In our case of heterogeneous space, using the above procedure and considering the left-reflecting defect in medium-1 at site $m = 1$ with $b=q_1(1+g_1)/2$, we obtain the left-bounded propagator generating function from Eq.~\eqref{eq:Defect-1-LB} as
\begin{align}
L^{\mathrm{(r)}}_{\mu,\nu}(n,z|n_0) &=  S_{\mu,\nu}(n,z|n_0) \nonumber \\
&+   \frac{z q_1 (1+g_1) \,  S_{1,\nu}(1,z|n_0) \big[   S_{\mu,1}(n,z|1) -  S_{\mu,1}(n,z|0)   \big]  }{2-z q_1 (1+g_1) \big[  S_{1,1}(1,z|1) - S_{1,1}(1,z|0)   \big]} \, ; \quad n,n_0 \geq 1 \, . \label{eq:Defect-1-LB-inhomo}
\end{align} 
Note that in writing Eq.~\eqref{eq:Defect-1-LB-inhomo}, the subscripts to the functions $L^{\mathrm{(r)}}(n,z|n_0)$ and $S(n,z|n_0)$ are not necessarily needed, they are present for later convenience to keep track of the embedding media of $n, n_0$; once $n, n_0$ are specified one may readily use the general solutions (Eqs.~\eqref{eq:GF_inZ_2reg_gen_sol} and~\eqref{eq:GF_inZ_tB_gen_sol}) for both \tA and \tB dynamics.
Now, using $L^{\mathrm{(r)}}_{\mu,\nu}(n,z|n_0)$ from Eq.~\eqref{eq:Defect-1-LB-inhomo} as the defect-free function and considering the right-reflecting defect in medium-2 at site $m = N$ with $c = q_2 (1-g_2)/2$, we obtain the fully-bounded propagator generating function from Eq.~\eqref{eq:Defect-1-RB} as
\begin{align}
& F^{\mathrm{(r)}}_{\mu,\nu}(n,z|n_0) =  L^{\mathrm{(r)}}_{\mu,\nu}(n,z|n_0)  \nonumber \\
&\hskip1cm +   \frac{z q_2 (1-g_2) \,  L^{\mathrm{(r)}}_{2,\nu}(N,z|n_0) \big[  L^{\mathrm{(r)}}_{\mu,2}(n,z|N)  - L^{\mathrm{(r)}}_{\mu,2}(n,z|N + 1)  \big]  }{2-z q_2 (1-g_2) \big[  L^{\mathrm{(r)}}_{2,2}(N,z|N)  - L^{\mathrm{(r)}}_{2,2}(N,z|N+1)  \big]}  ; \quad  n,n_0 \in [1, N]  . \label{eq:Defect-1-FB-inhomo}
\end{align}

The steady-state probability in the reflecting domain is defined by ${}^{\mathrm{ss}}F^{\mathrm{(r)}}_{\mu, \nu}(n|n_0) \equiv  \lim_{z\to 1} (1-z) F^{\mathrm{(r)}}_{\mu, \nu}(n,z|n_0) $, which reduces in the absence of bias (as $g_1=g_2=0$) to (see~\ref{app:bounded-steady-state})
\begin{align}
{}^{\mathrm{ss}}F^{\mathrm{(r)}}_{\mu, \nu}(n|n_0) =
\begin{cases} 
      \dfrac{1}{q_\mu \left( \frac{M}{q_1} + \frac{ N-M}{q_2} \right)} \, , & {\text{\tA}} \, ,\\[4ex]
      \dfrac{1}{N} \, , & {\text{\tB}}  \, .
\end{cases} \label{eq:ssTAB}
\end{align}
Note that, as expected, the quantity ${}^{\mathrm{ss}}F^{\mathrm{(r)}}_{\mu, \nu}(n|n_0)$ has no $n_0$-dependence as the index $\nu$ is absent on the right hand side of Eq.~\eqref{eq:ssTAB}, while the $n$-dependence appears through the index $\mu$ in $q_\mu$. 
Equation~\eqref{eq:ssTAB} clearly shows that the steady-sate probability without bias for \tA interface, despite being uniform within each medium, is discontinuous across the interface, thus displaying a step-like behavior over the full spatial range. The steady-state occupation probability of the walker in medium-1 and 2 is given by $M q_2/ [M q2+(N-M)q_1]$, and $(N-M) q_1/ [M q2+(N-M)q_1]$, respectively. On the other hand, for \tB interface, the steady-state probability without bias shows no discontinuity and is uniform throughout the heterogeneous space.

\begin{figure}[!htbp]
\centering
\includegraphics[scale=0.935]{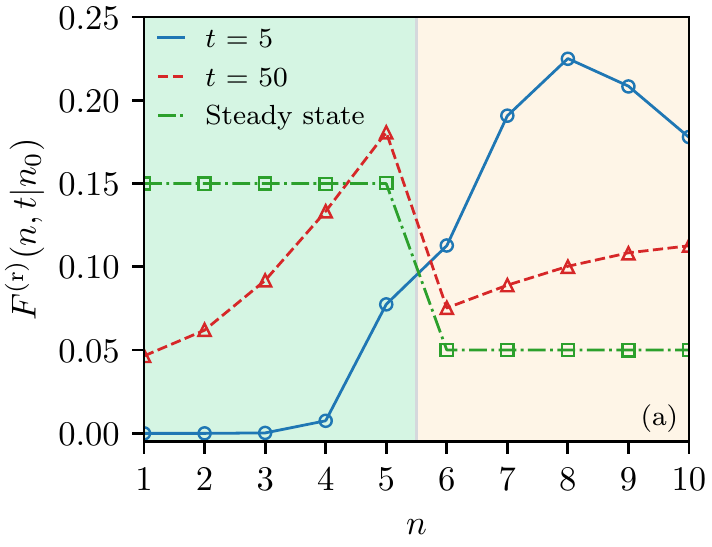}
\includegraphics[scale=0.935]{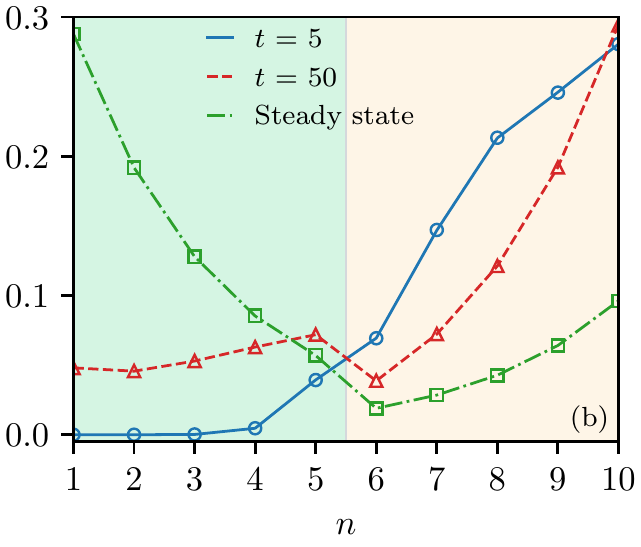}
\caption{Propagator $F^{\mathrm{(r)}}(n,t|n_0)$ in reflecting domain with \tA interface between sites 5 and 6. The walker starts from site $n_0=8$ in medium-2 in the domain of size $N=10$ with two different diffusivities $q_1=0.2$ and $q_2 = 0.6$. Without bias in panel (a), and with bias $g_1=0.2,\, g_2=-0.2$ in panel (b).  The lines are obtained by numerically inverting~\cite{abate_numerical_1992} Eq.~\eqref{eq:Defect-1-FB-inhomo} using Eqs.~\eqref{eq:Defect-1-LB-inhomo} and~\eqref{eq:GF_inZ_2reg_gen_sol}, while the points are obtained from $5\times 10^6$ stochastic realizations. }
\label{fig:typeA-2M-Bound-prop}
\end{figure}
\begin{figure}[!htbp]
\centering
\includegraphics[scale=0.935]{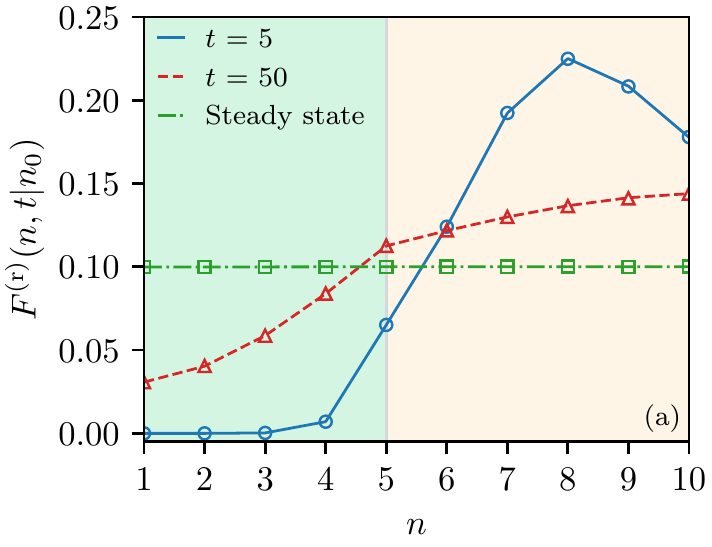}
\includegraphics[scale=0.935]{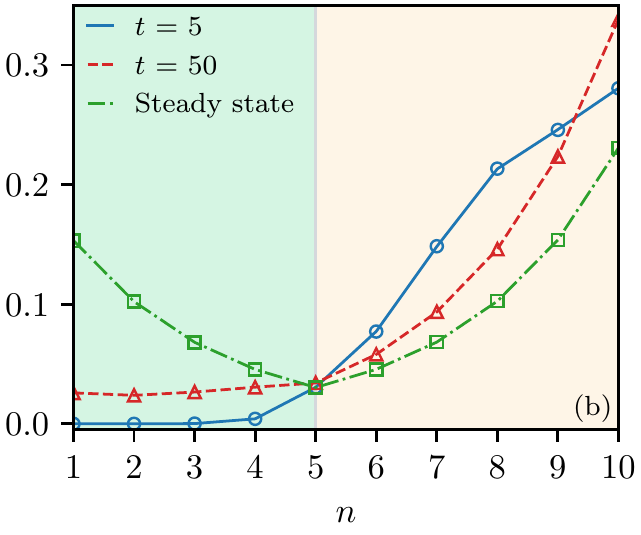}
\caption{Propagator $F^{\mathrm{(r)}}(n,t|n_0)$ in reflecting domain with \tB interface at site 5: (a) without bias, and (b) with bias. Parameter values are the same as in Fig.~\ref{fig:typeA-2M-Bound-prop}. The lines are obtained by numerically inverting~\cite{abate_numerical_1992} Eq.~\eqref{eq:Defect-1-FB-inhomo} using Eqs.~\eqref{eq:Defect-1-LB-inhomo} and~\eqref{eq:GF_inZ_tB_gen_sol}, while the points are obtained from $5\times 10^6$ stochastic realizations. }
\label{fig:typeB-2M-Bound-prop}
\end{figure}

Figures~\ref{fig:typeA-2M-Bound-prop}\, and~\ref{fig:typeB-2M-Bound-prop}\, depict the fully-bounded time-dependent propagator $F^{\mathrm{(r)}}(n,t|n_0)$ in reflecting domain with \tA and \tB interfaces, respectively. 
From Fig.~\ref{fig:typeA-2M-Bound-prop}(a), we see that in the absence of bias the discontinuity of the \tA propagator around the interface becomes prominent as time increases, and eventually it settles to a 
step-like steady state, i.e., it is separately uniform in each of the media with a jump at the interface. 
The ratio of the steady-state probability in medium-1 to that in medium-2 is $q_2/q_1$, see Eq.~\eqref{eq:ssTAB}.
On the other hand, the \tB propagator in the absence of bias, depicted in Fig~\ref{fig:typeB-2M-Bound-prop}(a), does not show such discontinuity as time increases.
In the steady-state, it rather becomes uniform throughout the heterogeneous space including the interface as suggested by Eq.~\eqref{eq:ssTAB}.
Figures~\ref{fig:typeA-2M-Bound-prop}(b)\, and~\ref{fig:typeB-2M-Bound-prop}(b) show the propagators with a symmetric and opposite bias away from the interface. 
As time increases, the \tA propagator in Fig.~\ref{fig:typeA-2M-Bound-prop}(b) shows discontinuity around the interface that also persists in the steady state.
Such discontinuity is once again not seen for the \tB propagator in Fig.~\ref{fig:typeB-2M-Bound-prop}(b). 
As the magnitude of the biases in medium-1 and 2 are equal, the steady-state probability in this case becomes symmetric about the interface.
In both cases, as the biases are opposite and away from the interface, the steady-state probabilities are greater at the reflecting boundaries than that at the bulk. 

\subsection{Absorbing boundaries}
\label{subsec:abs-prop}

Let us now consider the heterogeneous space limited with two perfectly absorbing boundaries at sites $1$ and $N$. 
One may again employ the two-step single-defect technique to obtain the fully-bounded propagator in this absorbing domain.
A different procedure that makes use of a determinant calculation when there exist multiple partially absorbing sites can be found in Ref.~\cite{giuggioli_spatio-temporal_2022}.
Considering the left defective site $1$ as the single perfectly absorbing point, one obtains the left-bounded propagator as
\begin{align}
L^{(\mathrm{a})}_{\mu,\nu}(n,z|n_0) =  S_{\mu,\nu}(n,z|n_0) - \frac{S_{\mu,1}(n,z|1) \, S_{1,\nu}(1,z|n_0) }{S_{1,1}(1,z|1)} \, ; \quad n,n_0 \geq 1 \, , \label{eq:Defect-1-LB-inhomo-abso}
\end{align} 
where once again the subscripts to the functions $L^{(\mathrm{a})}(n,z|n_0)$ and $S(n,z|n_0)$ are kept only for convenience, while the superscript $a$ represents the absorbing boundaries.
In this case, because of the absorbing boundary, the quantity $L^{(\mathrm{a})}_{\mu,\nu}(n,z|n_0)$ is not normalized.
However, as the normalization is not required to apply the defect technique, one may safely run another defect iteration on it. 
Now, starting with $L^{(\mathrm{a})}_{\mu,\nu}(n,z|n_0)$ as the defect-free propagator and considering the right absorbing site $N$, the fully-bounded propagator is obtained as
\begin{align}
F^{(\mathrm{a})}_{\mu,\nu}(n,z|n_0) =  L^{(\mathrm{a})}_{\mu,\nu}(n,z|n_0) - \frac{L^{(\mathrm{a})}_{\mu,2}(n,z|N) \, L^{(\mathrm{a})}_{2,\nu}(N,z|n_0) }{L^{(\mathrm{a})}_{2,2}(N,z|N)} \, ; \quad n,n_0 \in [1, N] \, . \label{eq:Defect-1-FB-inhomo-abso}
\end{align}

\subsection{Mixed boundaries}
\label{subsec:mix-prop}

In case the space is confined with mixed boundaries, e.g., a reflecting boundary at site 1 and an absorbing boundary at site $N$, the fully-bounded propagator is given by
\begin{align}
F^{(\mathrm{m})}_{\mu,\nu}(n,z|n_0) =  L^{(\mathrm{r})}_{\mu,\nu}(n,z|n_0) - \frac{ L^{(\mathrm{r})}_{\mu,2}(n,z|N)   \,   L^{(\mathrm{r})}_{2,\nu}(N,z|n_0)}{L^{(\mathrm{r})}_{2,2}(N,z|N)} \, ; \quad n,n_0 \in [1, N] \, , \label{eq:Defect-1-FB-inhomo-mixed}
\end{align}
where $L^{(\mathrm{r})}_{\mu,\nu}(n,z|n_0)$ is the left-bounded reflecting propagator given in Eq.~\eqref{eq:Defect-1-LB-inhomo}.

\section{First-passage statistics in a reflecting domain}
\label{subsec:fp}

In a reflecting domain, the first-passage probability from site $n_0$ in medium-$\nu$ to site $n$ in medium-$\mu$ in the $z$-domain is given by
\begin{align}
\label{eq:fp-TAB}
\mathcal{F}^{\mathrm{(r)}}_{\mu,\nu}(n,z|n_0) \equiv \frac{F^{\mathrm{(r)}}_{\mu, \nu}(n,z|n_0) }{F^{\mathrm{(r)}}_{\mu, \mu}(n,z|n) } = \frac{  \mathbb{C}(z) L^{\mathrm{(r)}}_{\mu, \nu}(n,z|n_0)   + \mathbb{M}_{\mu}(n,z) \, \mathbb{R}_{\nu}(n_0,z) }{  \mathbb{C}(z) L^{\mathrm{(r)}}_{\mu, \mu}(n,z|n)   + \mathbb{M}_{\mu}(n,z) \, \mathbb{R}_{\mu}(n,z)} \, ,
\end{align}
where the functions $\mathbb{M}_{\mu}(n,z)$, $\mathbb{C}(z)$, and $\mathbb{R}_{\nu}(n_0,z)$ are defined in Eqs.~\eqref{eq:def-M-mu-n-z},~\eqref{eq:def-C-z}, and~\eqref{eq:def-R-nu-n0-z}, respectively (see~\ref{app:bounded-steady-state}).
In the absence of bias, the mean first-passage time (MFPT) to reach site $n$ in medium-$\mu$ from site $n_0$ in medium-$\nu$ defined by $T^{\mathrm{(r)}}_{\substack{{n_0 \to n}\\{\nu \to \mu}}} \equiv \lim_{z \to 1} \pdv{z}\mathcal{F}^{\mathrm{(r)}}_{\mu,\nu}(n,z|n_0)$ is obtained as (see~\ref{app:bounded-FP})
\begin{align}
T^{\mathrm{(r)}}_{\substack{{n_0 \to n}\\{1 \to 1}}} &= \dfrac{(n-n_0)(n+n_0-1)}{q_1} - \dfrac{n-n_0-|n-n_0|}{q_1 q_2} \big[  M q_2 + (N-M) q_1 \big] \, ,  \label{eq:MFPT-refl-TA1}\\
T^{\mathrm{(r)}}_{\substack{{n_0 \to n}\\{1 \to 2}}} &=    \dfrac{(n-M)(n-M-1)}{q_2} + \dfrac{M (2 n-M-1) - n_0^2+n_0}{q_1} \, ,  \label{eq:MFPT-refl-TA2}\\
T^{\mathrm{(r)}}_{\substack{{n_0 \to n}\\{2 \to 1}}} &=    \dfrac{(n-M)(n-M-1)}{q_1} + \dfrac{M (2 n-M-1) - n_0^2+n_0-2 N (n-n_0)}{q_2}  \, , \label{eq:MFPT-refl-TA3}\\
T^{\mathrm{(r)}}_{\substack{{n_0 \to n}\\{2 \to 2}}} &=    \dfrac{(n-n_0)(n+n_0-1)}{q_2} + \dfrac{(n-n_0) \big[ M (q_1-q_2) (q_1-2q_2) - N q_1^2 \big]}{q_1 q_2^2} \nonumber \\
& + \dfrac{|n-n_0|}{q_2^2} \big[  M q_2 + (N-M) q_1 \big]  \, , \label{eq:MFPT-refl-TA4}
\end{align}
for \tA dynamics, and
\begin{align}
\label{eq:MFPT-refl-TB}
T^{\mathrm{(r)}}_{\substack{{n_0 \to n}\\{\nu \to \mu}}} &= 
\begin{cases}
\dfrac{(n-n_0)(n+n_0-1-N)+ |n-n_0|N}{q_\nu} \, , &   \quad   \begin{cases} \nu=1, & \mu =1,c,  \\ \nu=2, & \mu=2,c, \end{cases} \\
\dfrac{(n-M)(n+M-1-N)+ |n-M|N}{q_\mu} \, , & \quad   \nu=c, \quad \mu = 1,2, \\
T^{\mathrm{(r)}}_{\nu \to c} + T^{\mathrm{(r)}}_{c \to \mu} \, ,& \quad  \begin{cases} \nu=1, & \mu =2,  \\ \nu=2, & \mu=1, \end{cases}
\end{cases}
\end{align}
for \tB dynamics.

Striking features of the MFPT from $n_0$ to target $n$ in the heterogeneous space may be seen from the above equations.
For the \tB dynamics, when the target $n$ and $n_0$ are located in the same medium (i.e. for $\nu=\mu$) or the interface itself is the target, Eq.~\eqref{eq:MFPT-refl-TB} clearly suggests that the underlying heterogeneity has no effect on the MFPT. 
In this case, the MFPT becomes equal to that in a homogeneous space of size $N$ confined in a reflecting domain with diffusivity $q_\nu$~\cite{giuggioli_exact_2020}. 
When the walker starts from the interface at $M$, the MFPT depends only on the diffusivity of the target medium-$\mu$ in the same way as if the whole space of size $N$ becomes homogeneous with diffusivity $q_\mu$.
In the particular case, when the target medium is different from the source medium, the MFPT is given by the sum of two MFPTs: (i) from the source to the interface, (ii) from the interface to the target.
On the other hand, for the \tA dynamics, the heterogeneity of space prevails in all possible scenarios because the MFPT (Eq.~\eqref{eq:MFPT-refl-TA1}--\eqref{eq:MFPT-refl-TA4}) evidently depends on the diffusivity of both media. This persists because of the mixing of the two diffusivities $q_1$ and $q_2$ in the jump probabilities between sites $M$ and $M+1$ at the \tA interface (see Fig~\ref{fig:tA-dynamics}). 
Such mixing is absent at the \tB interface because the two diffusivities only modifies the probability of staying at the interface (see Fig~\ref{fig:tB-dynamics}).

\section{Continuous limit}
\label{sec:conti}

\subsection{Propagator and boundary conditions at the interface}
\label{sec:conti-pbdy}

To obtain the propagator in the continuous space-time limit, we start by denoting the waiting time distribution, i.e., the probability time density which is required to perform a single hop, by $\phi(\tau)$, and its Laplace transform by  $\widetilde{\phi}(\epsilon) \equiv \int_{0}^{\infty} \dd{\tau} \exp({-\epsilon \tau}) \, \phi(\tau) $ with $\tau $ being the continuous time variable. The discrete-space continuous-time propagator in the Laplace domain $\widetilde{P}_{\mu,\nu}(n,\epsilon|n_0)$ is related to the discrete-space discrete-time propagator generating function $S_{\mu,\nu}(n,z|n_0)$ by the general formula~\cite{montroll_random_1965} $\widetilde{P}_{\mu,\nu}(n,\epsilon|n_0) = [({1- \widetilde\phi(\epsilon)})/{\epsilon}] \,S_{\mu,\nu} \big(n,z \to \widetilde{\phi}(\epsilon) |n_0 \big)$,
which for an exponential waiting time distribution $\phi(\tau) = R \exp(-R\tau)$, with $R$ being the hopping rate, reduces to 
\begin{align}
\widetilde{P}_{\mu,\nu}(n,\epsilon|n_0) = \frac{S_{\mu,\nu} \big(n,z \to \frac{R}{\epsilon + R} |n_0 \big)}{\epsilon +R} \, .  \label{eq:DSCT-gen}
\end{align} 
Denoting the lattice spacing by $a$, the continuous-space continuous-time (CSCT) propagator in the Laplace domain, $\widetilde{P}_{\mu,\nu}(x,\epsilon|x_0)$, is obtained by using the relation $\widetilde{P}_{\mu,\nu}(x,\epsilon|x_0) \equiv \widetilde{P}_{\mu,\nu}(n,\epsilon|n_0) / a$, and taking the simultaneous limits
$a \to 0$, $R \to \infty$, $n \to \infty$, $ g_\mu \to 0 $ such that $x \equiv n a$, $x_0 \equiv n_0 a$, $D_\mu \equiv q_\mu R a^2 / 2$, and $\gamma_\mu \equiv q_\mu g_\mu R a$. Here, $D_\mu$ is the diffusion constant in medium-$\mu$, while $\gamma_\mu$ is the strength of the bias in units of velocity in the same medium. By considering these limits and choosing $x_0 \in $ medium-$1$, the CSCT propagator in the Laplace domain is obtained using Eqs.~\eqref{eq:GF_inZ_1_sol},~\eqref{eq:GF_inZ_2_sol},~\eqref{eq:GF_inZ_S1_sol_V},~\eqref{eq:GF_inZ_S2_sol_V}, and~\eqref{eq:DSCT-gen}, as 
\begin{align}
\label{eq:CSCT-11}
&\widetilde{P}_{1,1}(x,\epsilon|x_0) = \frac{\ee^{-(x-x_0)\frac{\gamma_1}{2 D_1}}}{\sqrt{\gamma^2_1 + 4 \epsilon D_1}} \Bigg[ \exp\Big\{-\frac{|x-x_0|}{2 D_1} \sqrt{\gamma^2_1 + 4 \epsilon D_1} \Big\} \nonumber \\
& -  \frac{ \rho \big(\gamma_1 -\sqrt{\gamma^2_1 + 4\epsilon D_1}  \big) -  \gamma_2 +\sqrt{\gamma^2_2 + 4\epsilon D_2}   }{ \rho \big(\gamma_1 +\sqrt{\gamma^2_1 + 4\epsilon D_1}  \big) -  \gamma_2 +\sqrt{\gamma^2_2 + 4\epsilon D_2}  }  \exp\Big\{{- \frac{ (2 x_M -x -x_0)}{2 D_1} \sqrt{\gamma^2_1 + 4 \epsilon D_1}  }\Big\}  \Bigg]  ,
\end{align}
\begin{align}
\label{eq:CSCT-21}
&\widetilde{P}_{2,1}(x,\epsilon|x_0) = \frac{2  \exp \Big[{- \frac{(x_M-x_0)}{2 D_1}(\gamma_1 + \sqrt{\gamma^2_1 + 4 \epsilon D_1}) - \frac{(x-x_M)}{2 D_2}(\gamma_2 + \sqrt{\gamma^2_2 + 4 \epsilon D_2}) \Big] }}{  \rho \big(\gamma_1 +\sqrt{\gamma^2_1 + 4\epsilon D_1} \, \big) -  \gamma_2 +\sqrt{\gamma^2_2 + 4\epsilon D_2}  } \, ,
\end{align}
where $x_M$ is the location of the interface between the two media, and
\begin{align}
\rho = 
\begin{cases} 
      \dfrac{D_2}{D_1} \, , & {\text{\tA}} \, ,\\[2ex]
      1 \, , & {\text{\tB}}  \, .
\end{cases} \label{eq:rho-def}
\end{align}
For $x_0 \in $ medium-$2$, the CSCT propagators $\widetilde{P}_{1,2}(x,\epsilon|x_0)$, $\widetilde{P}_{2,2}(x,\epsilon|x_0)$ obtained in the same limits are given in Eqs.~\eqref{eq:CSCT-12}--\eqref{eq:CSCT-22} in~\ref{app:CSCT-Reg2}.

From Eqs.~\eqref{eq:CSCT-11} and~\eqref{eq:CSCT-21}, one may straightforwardly show that at $x=x_M$ the functions $\widetilde{P}_{1,1}(x,\epsilon|x_0)$ and $\widetilde{P}_{2,1}(x,\epsilon|x_0)$ satisfy the following boundary conditions
\begin{align}
& \widetilde{P}_{1,1}(x_M,\epsilon|x_0) = \rho \, \widetilde{P}_{2,1}(x_M,\epsilon|x_0) \, , \label{eq:p-CSCT-bdy1} \\
& \widetilde{\mathcal{J}}_{1,1}(x_M,\epsilon|x_0) =  \widetilde{\mathcal{J}}_{2,1}(x_M,\epsilon|x_0) \, , \label{eq:p-CSCT-bdy2}
\end{align}
with the current defined as
\begin{align}
& \widetilde{\mathcal{J}}_{\mu,\nu}(x,\epsilon|x_0) \equiv -\gamma_\mu \widetilde{P}_{\mu,\nu}(x,\epsilon|x_0) -D_\mu \frac{\partial \widetilde{P}_{\mu,\nu}(x,\epsilon|x_0)}{\partial x}  \, . \label{eqL:J-def}
\end{align}
For $x_0 \in $ medium-$2$, Eqs.~\eqref{eq:CSCT-12} and~\eqref{eq:CSCT-22} yield the same boundary conditions, namely, $\widetilde{P}_{1,2}(x_M,\epsilon|x_0) = \rho \, \widetilde{P}_{2,2}(x_M,\epsilon|x_0)$ and $\widetilde{\mathcal{J}}_{1,2}(x_M,\epsilon|x_0) =  \widetilde{\mathcal{J}}_{2,2}(x_M,\epsilon|x_0)$.
The first boundary condition~\eqref{eq:p-CSCT-bdy1} clearly shows that in heterogeneous space ($D_1\neq D_2$), the probability density at a \tA interface is discontinuous, while at a \tB interface it is continuous. The second boundary condition~\eqref{eq:p-CSCT-bdy2} shows the equality of fluxes at both \tA and \tB interfaces and warrants the fact that there is no accumulation at the $x=x_M$.  
We note that the obtained \tA and the \tB boundary conditions are, respectively, equivalent to the It\^{o} and the post-point (or anti-It\^{o} or H\"{a}nggi-Klimontovich) interpretations of stochastic differential equations at the interface in continuous space and time~\cite{vaccario_first-passage_2015}.

In homogeneous continuous space ($D_\mu=D$) with the same bias $\gamma_\mu =\gamma$, the quantities in Eqs.~\eqref{eq:CSCT-11} and~\eqref{eq:CSCT-21} may be combined into a single expression
\begin{align}
\widetilde{P}(x,\epsilon|x_0) = \frac{1}{\sqrt{\gamma^2 + 4 \epsilon D}} ~ \exp\bigg[ {- \frac{(x-x_0)\gamma}{2D}- \frac{ |x-x_0|}{2 D} \sqrt{\gamma^2 + 4 \epsilon D}} \bigg] \, ,
\end{align}
which upon an inverse Laplace transform yields the known solution with a bias towards negative $x$-axis, namely, $P(x,\tau|x_0) = \exp[-(x-x_0+\gamma \tau)^2/(4 D \tau)]/\sqrt{4\pi D \tau}$. 
Closed-form expressions for the time-dependent propagator are obtained for heterogeneous space with $D_1\neq D_2$ and $\gamma_\mu=0$, and given in Eqs.~\eqref{eq:CSCT-11-g0-int}--\eqref{eq:CSCT-21-g0-int} in~\ref{app:CSCT-Reg2}.

\subsection{Leather boundary condition at a permeable barrier}
\label{sec:leather}

In solving the diffusion equation $ \partial_\tau {P}(x, \tau) = \gamma {P}(x, \tau) + D \, \partial^2_x {P}(x, \tau)$ with a global bias $\gamma$ in a homogeneous space with a partially permeable barrier at $x=x_b$, a commonly used boundary condition in the literature is the so-called leather boundary condition or permeable boundary condition (PBC)~\cite{tanner_transient_1978,powles_exact_1992},
\begin{align}
\mathcal{J}(x^{\pm}_b,\tau) = \kappa [ P(x_b^{-},\tau) - P(x_b^+,\tau)] \, , \label{eq:leather-bc}
\end{align}
where $\mathcal{J}(x,\tau) \equiv - \gamma P(x,\tau) - D  \partial_x P(x,\tau)$ is the probability current, $\kappa$ is the permeability of the barrier, and the $\pm$ superscript denotes the respective sides of the barrier. 
We show now how the PBC can be derived in the heterogeneous setting by considering diffusion dynamics in a space of three different media  separated by \tB interfaces, and squeezing the medium in the middle to obtain the permeable barrier.
For that we consider that the full heterogeneous space is separated into three media by two \tB interfaces at $x_1$ and $x_2$ such that the spatial regions $x<x_1$, $x_1< x < x_2$, and $x>x_2$ belong to medium-1, 2, and 3 having diffusion constants $D_1$, $D_2$, and $D_3$, respectively.
The biases in medium-1 and 3 are given by $\gamma_1$, and $\gamma_3$, respectively, while we assume that there is no bias in medium-2, i.e., $\gamma_2=0$.
To solve the diffusion problem one may choose a localized initial condition $P(x,0)=\delta(x-x_0)$ where $x_0 \in$ medium-1.
Because of the initial condition, the full domain of $x$ naturally breaks up into four regions separated by three points $x_0$, $x_1$, and $x_2$~\cite{chase_analysis_2016}.

In the Laplace domain, we denote the propagators by $\widetilde{P}^{(\mathrm{I})}_{1,1}(x, \epsilon)$, $\widetilde{P}^{(\mathrm{II})}_{1,1}(x, \epsilon)$, $\widetilde{P}_{2,1}(x, \epsilon)$, and $\widetilde{P}_{3,1}(x, \epsilon)$ for the four regions  $x<x_0$, $x_0 < x<x_1$, $x_1<x<x_2$, and $x>x_2$, respectively. This problem may be solved separately in each of the four regions, wherein it reduces to a homogeneous constant-coefficient differential equation of the form $D_\mu \, \partial^2_x \widetilde{P}_{\mu,\nu}(x, \epsilon) + \gamma_\mu \partial_x  \widetilde{P}_{\mu,\nu}(x, \epsilon) - \epsilon \, \widetilde{P}_{\mu,\nu}(x, \epsilon) = 0$ that has an exponential solution $\widetilde{P}(x,\epsilon) = A \exp(x \alpha_{-}) + B \exp(- x \alpha_{+})$ 
where $\alpha_{\pm} \equiv ( \pm \gamma_\mu + \sqrt{\gamma^2_\mu + 4 \epsilon D_\mu} )/(2 D_\mu)$ with $A$ and $B$ being two arbitrary constants~\cite{chase_analysis_2016,redner_guide_2001}.
The global solution is obtained by joining the solutions in the four regions and computing eight arbitrary constants using the boundary conditions.
The requirement that $\widetilde{P}^{(\mathrm{I})}_{1,1}(x,\epsilon)$ and $\widetilde{P}_{3,1}(x,\epsilon)$ must vanish as $x \to -\infty$ and $+\infty$, respectively, eliminates two of the constants. 
At $x=x_0$, the continuity condition $\widetilde{P}^{(\mathrm{I})}_{1,1}(x_0,\epsilon) = \widetilde{P}^{(\mathrm{II})}_{1,1}(x_0,\epsilon)$ and the jump condition $ \partial_x \widetilde{P}^{(\mathrm{II})}_{1,1}(x,\epsilon)|_{x_0} - \partial_x\widetilde{P}^{(\mathrm{I})}_{1,1}(x,\epsilon)|_{x_0}  =-1/D_1$ determine two more constants~\cite{redner_guide_2001}. 
The remaining four constants are obtained using the \tB boundary conditions (Eqs.~\eqref{eq:p-CSCT-bdy1}--\eqref{eq:p-CSCT-bdy2}) at $x=x_1$ and $x=x_2$, namely, $\widetilde{P}^{(\mathrm{II})}_{1,1}(x_1,\epsilon) = \widetilde{P}_{2,1}(x_1,\epsilon)$, $\widetilde{\mathcal{J}}^{(\mathrm{II})}_{1,1}(x_1,\epsilon) = \widetilde{\mathcal{J}}_{2,1}(x_1,\epsilon)$, $\widetilde{P}_{2,1}(x_2,\epsilon) = \widetilde{P}_{3,1}(x_2,\epsilon)$, and $\widetilde{\mathcal{J}}_{2,1}(x_2,\epsilon) = \widetilde{\mathcal{J}}_{3,1}(x_2,\epsilon)$. 
We then squeeze medium-2 by taking the limit $\delta \equiv ( x_2-x_1) \to 0$ so that medium-1 and 3 become separated by a partially permeable barrier at $x_1$ with permeability $\kappa \equiv D_2/\delta$. In this limit, we obtain the boundary condition
\begin{align}
\widetilde{\mathcal{J}}^{(\mathrm{II})}_{1,1}(x_1,\epsilon) = \widetilde{\mathcal{J}}_{3,1}(x_1,\epsilon) = \kappa \left[ \widetilde{P}^{(\mathrm{II})}_{1,1}(x_1,\epsilon) - \widetilde{P}_{3,1}(x_1,\epsilon) \right]   \, .\label{eq:my-leather}
\end{align}
to be satisfied at the permeable barrier.
Note that the boundary condition~\eqref{eq:my-leather} is the generalization of the PBC~\eqref{eq:leather-bc} in heterogeneous space, where two media having different diffusion constants ($D_1$, $D_3$) and biases ($\gamma_1$, $\gamma_3$) are separated by a permeable barrier at $x_1$.

\section{Conclusion}
\label{sec:conclu}

In this paper, we have studied the DSDT dynamics of a random walker moving in a heterogeneous space of two media separated by an interface.
We have shown two exclusive ways to model the interface depending upon whether it is placed between two lattice sites (\tA) or on a lattice site shared by both the media (\tB).
For both cases, we have obtained exact results on the propagator in unbounded domain. 
At the interface, the \tA propagator shows a discontinuous behavior, while the \tB propagator is smoother.
We have also demonstrated how to use the defect technique to obtain the propagator in domains confined by reflecting, absorbing and mixed boundaries. 
It is shown that the steady-state probability in a reflecting domain without bias shows a step-like uniform behavior with a jump at the interface for \tA dynamics, while it is uniform throughout the heterogeneous space for the \tB dynamics.
For both dynamics, we have also obtained the first-passage probability and the mean first-passage time in the reflecting domain without bias.
Taking the CSCT limit on the DSDT propagator, we have obtained the boundary conditions~\eqref{eq:p-CSCT-bdy1} and~\eqref{eq:p-CSCT-bdy2} at the interface. The first boundary condition shows that the probability density at the interface is discontinuous for \tA dynamics, but continuous for \tB dynamics. The second boundary condition shows that the flux at the interface from both media are the same for both the dynamics. 
Finally, we have derived a generalized boundary condition in presence of a permeable barrier.

As an extension to this work, one may study the transmission dynamics between randomly moving individuals in a heterogeneous space. 
Random transmission of information between moving agents in short scales often controls patterns emerging at large scales.
An analytic framework to study such transmission dynamics in arbitrary spatial domains has been derived recently~\cite{giuggioli_spatio-temporal_2022}. 
Within this framework, one may make use of the propagators given in Eqs.~\eqref{eq:GF_inZ_2reg_gen_sol} and~\eqref{eq:GF_inZ_tB_gen_sol} to study transmissions in heterogeneous space.
It would be also interesting to study the effects of spatial heterogeneity on the biased random walk dynamics in presence of stochastic resetting~\cite{ evans_diffusion_2011}. A discrete renewal equation in presence of stochastic resetting has been derived in Ref.~\cite{das_discrete_2022}, where working formula are presented to compute different quantities in terms of the reset-free propagator. To study resetting in heterogeneous space, one may use these formula with the propagators derived here.
Another important future direction would be to study the heterogeneous dynamics in higher dimensions, one way of doing which, could be the hierarchical dimensionality reduction procedure~\cite{giuggioli_exact_2020}.

\section{Acknowledgments}
\label{sec:ackno}

We acknowledge funding from the Biotechnology and Biological Sciences Research Council (BBSRC) Grant No. BB/T012196/1. We thank Seeralan Sarvaharman and Toby Kay for helpful discussions.

\appendix
\addtocontents{toc}{\fixappendix}

\section{Solving the \tA Master equation}
\label{app:tA-solutions}

For an initial site $n_0 \in$ medium-1, i.e., for $n_0 \leq M$, the \tA Master equation~\eqref{eq:master_eq_1}--\eqref{eq:master_eq_4} in the $z$-domain are given by 
\begin{align}
 S_{1,1} (n,z|n_0) \left[ 1 -  z(1-q_1) \right] &= \delta_{n,n_0} + \frac{z q_1}{2} (1-g_1)  S_{1,1}(n-1,z|n_0) \nonumber \\ 
& + \frac{z q_1}{2} (1+g_1) S_{1,1}(n+1,z|n_0)   \, ; \quad n < M  \, , \label{eq:master_eq_inZ_1} \\
 S_{1,1} (M,z|n_0) \left[ 1 -  z(1-q_1) \right] &= \delta_{M,n_0} + \frac{z q_1}{2}  (1-g_1) S_{1,1}(M-1,z|n_0)  \nonumber \\ 
& +  \frac{z q_2}{2} (1+g_2) S_{2,1}(M+1,z|n_0)    \, , \label{eq:master_eq_inZ_2} \\
 S_{2,1} (M+1,z|n_0) \left[ 1 -  z(1-q_2) \right] &=  \frac{z q_1}{2} (1-g_1) S_{1,1}(M,z|n_0)  \nonumber \\ 
& + \frac{z q_2}{2} (1+g_2) S_{2,1}(M+2,z|n_0)  \, , \label{eq:master_eq_inZ_3} \\
 S_{2,1} (n,z|n_0) \left[ 1 -  z(1-q_2) \right] &=  \frac{z q_2}{2} (1-g_2) S_{2,1}(n-1,z|n_0) \nonumber \\ 
& + \frac{z q_2}{2} (1+g_2) S_{2,1}(n+1,z|n_0)   \, ; \quad n > (M+1) \, . \label{eq:master_eq_inZ_4}
\end{align}
Substituting Eqs.~\eqref{eq:master_eq_inZ_1}--\eqref{eq:master_eq_inZ_4} in Eqs.~\eqref{eq:GF_inU_1_def}--\eqref{eq:GF_inU_2_def}, we obtain
\begin{align}
 G_{1,1} (u,z|n_0) \left[ 1 - \frac{\beta^{-}_1}{2} u - \frac{\beta^{+}_1}{2}    u^{-1} \right] &= \frac{u^{n_0}}{1 -z + z q_1} - \frac{\beta_1^{-} }{2} \, S_{1,1}(M,z|n_0) \, u^{M+1} \nonumber \\
& + \frac{ \beta_1^{-} q_2(1+g_2)   }{2 \, q_1(1-g_1)}  \,  S_{2,1}(M+1,z|n_0) \, u^{M} \, ,  \label{eq:GF_inU_1} \\
 G_{2,1} (u,z|n_0) \left[ 1 - \frac{\beta^{-}_2}{2} u - \frac{\beta^{+}_2}{2}    u^{-1} \right] &= \frac{ \beta_2^{+} q_1 (1-g_1)    }{2 \, q_2 (1+g_2)} \,  S_{1,1}(M,z|n_0) \, u^{M+1} \nonumber \\
& - \frac{\beta_2^{+} }{2} \, S_{2,1}(M+1,z|n_0) \, u^{M} \, , \label{eq:GF_inU_2}
\end{align}
where $\beta^{\pm}_\mu(z)$ is defined in Eq.~\eqref{eq:betaPM_i_def}.

Using the integral relation (see~\ref{app:contour}\,) 
\begin{align}
\frac{1}{2 \pi \ii} \oint_{C} {\mathrm d}u \, \dfrac{u^{n_0} }{u^{n+1} \left[ 1 - \dfrac{\beta^{-}_\mu}{2} u - \dfrac{\beta^{+}_\mu}{2}    u^{-1} \right] } = \frac{ f_\mu^{  \frac{ n - n_0 + |n-n_0|}{2} } \xi_\mu^{|n-n_0|} }{\sqrt{1-\beta^{+}_\mu \beta^{-}_\mu}} \, , \label{eq:Inversion_integral2}
\end{align}
where $f_\mu$ and $\xi_\mu(z)$ are defined in Eq.~\eqref{eq:f_i_def} and~\eqref{eq:xi_i_def}, respectively,
one may invert Eqs.~\eqref{eq:GF_inU_1}--\eqref{eq:GF_inU_2} according to Eq.~\eqref{eq:Inverse_Fourier_def_in_u} and obtain the generating functions $S_{1,1} (n,z|n_0)$ and $S_{2,1} (n,z|n_0)$ as
\begin{align}
 S_{1,1} (n,z|n_0)  &= \frac{1}{\sqrt{1-\beta^+_1 \beta^-_1}}  \Bigg[   \frac{f_1^{\frac{n-n_0+|n-n_0|}{2}} \xi_1^{|n-n_0|}}{1-z+z q_1} - \frac{\beta^-_1 }{2}  \, \xi_1^{M+1-n} S_{1,1}(M,z|n_0)  \nonumber \\
&\hskip10pt + \frac{\beta^-_1 q_2 (1+g_2) }{2 \,q_1 (1-g_1)}  \, \xi_1^{M-n}  S_{2,1}(M+1,z|n_0)   \Bigg] \, , \label{eq:GF_inZ_1} \\
 S_{2,1} (n,z|n_0)  &= \frac{1}{\sqrt{1-\beta^+_2 \beta^-_2}}  \Bigg[  \frac{\beta^+_2 q_1 (1-g_1) }{2\, q_2 (1+g_2)} \, (f_2 \xi_2)^{n-M-1} S_{1,1}(M,z|n_0) \nonumber \\
&\hskip10pt - \frac{\beta^+_2}{2}   (f_2\xi_2)^{n-M} S_{2,1}(M+1,z|n_0)  \Bigg]   \, . \label{eq:GF_inZ_2}
\end{align}
By putting $n=M$ in Eq.~\eqref{eq:GF_inZ_1} and $n=(M+1)$ in Eq.~\eqref{eq:GF_inZ_2}, and simultaneously solving the two equations, we obtain
\begin{align}
S_{2,1} (M+1,z|n_0) &= \frac{q_1(1-g_1)}{q_2(1+g_2)} \, \xi_2 \, S_{1,1}(M,z|n_0) \, , \\
S_{1,1} (M,z|n_0) &= \frac{2 \,(f_1\xi_1)^{M-n_0}}{ \big( 1-z+z q_1\big)  \big( 2 - \beta_1^- \xi_1 - \beta_1^- \xi_2 \big) } \, ,
\end{align}
which when substituted back in Eqs.~\eqref{eq:GF_inZ_1} and~\eqref{eq:GF_inZ_2} finally yield the solutions given in Eq.~\eqref{eq:GF_inZ_1_sol} and~\eqref{eq:GF_inZ_2_sol} of the main text.
For $n_0 \in$ medium-2, i.e., with $n_0 \geq (M+1) $, following the same procedure one gets
\begin{align}
 S_{1,2} (n,z|n_0) &= \frac{ 2 \, q_2 (1+g_2) \, \xi_1^{M+1-n} \, \xi_2^{n_0 - M-1} }{ q_1 (1+g_1) \big(1-z+z q_2\big) \big(2 - \beta^{+}_2 f_2 \xi_2  - \beta^{+}_2 f_1 \xi_1 \big) } \, , \label{eq:GF_inZ_1_sol_Reg2} \\
 S_{2,2} (n,z|n_0) &= \frac{1}{\sqrt{1-\beta^{+}_2 \beta^{-}_2} } \Bigg[  \frac{f_2^{\frac{n-n_0+|n-n_0|}{2}} \xi_2^{|n-n_0|}}{1-z+z q_2}  \nonumber \\
&\hskip10pt - \frac{\beta^{+}_2 \big( f_2 \xi_2  - f_1 \xi_1 \big)  f_2^{n-M-1} \, \xi_2^{n+n_0-2 M-2} }{\big(1-z+z q_2\big) \big(2 - \beta^{+}_2 f_2 \xi_2  - \beta^{+}_2 f_1 \xi_1 \big) }   \Bigg] \, . \label{eq:GF_inZ_2_sol_Reg2}
\end{align}

\section{Derivation of an integral relation}
\label{app:contour}

Let us compute the integral
\begin{align}
\label{eq:app-int-def}
I = \frac{1}{2 \pi \ii} \oint_C \dd{u} \frac{u^m}{u \left[ 1-\dfrac{\beta^{-}_\mu}{2} u -\dfrac{\beta^{+}_\mu}{2} u^{-1}\right]} \, ,
\end{align}
where $C$ is the counterclockwise unit circular contour centered at $u=0$ on the complex $u$-plane with $m \in \mathbb{Z}$ and $\beta^{\pm}_{\mu}$ is defined in Eq.~\eqref{eq:betaPM_i_def}. 
One may conveniently rewrite Eq.~\eqref{eq:app-int-def} as
\begin{align}
\label{eq:app-int-modi}
I = - \frac{2}{\beta^{-}_\mu} \frac{1}{2 \pi \ii} \oint_C \dd{u} \frac{u^m}{(u-u_{-}) (u-u_{+})} \, ,
\end{align}
where we have
\begin{align}
u_{-} = \frac{1 -  \sqrt{1-\beta^{+}_\mu\beta^{-}_\mu}}{\beta^{-}_\mu} \, ,\quad u_{+} =  \frac{1 +  \sqrt{1-\beta^{+}_\mu\beta^{-}_\mu}}{\beta^{-}_\mu} \, ,
\end{align}
with $|u_{-}| < 1 $ and $|u_{+}|>1$.

For $m\geq0$ the integrand in Eq.~\eqref{eq:app-int-modi} has two simple poles at points $u=u_{\pm}$, of which only the pole at $u=u_{-}$ lies inside the closed contour $C$. Using the residue theorem one may compute the integral to be
\begin{align}
I_{m\geq0} = -\frac{2}{\beta^{-}_\mu} \frac{u^{m}_{-}}{(u_{-} - u_{+})} = \frac{u^{m}_{-}}{\sqrt{1-\beta^{+}_\mu\beta^{-}_\mu}}  \label{eq:app-int-mleq0}
\end{align}
For $m<0$, substituting $m=-|m|$ and $s \equiv 1/u$, the integral~\eqref{eq:app-int-modi} may be recast as
\begin{align}
I_{m<0} &= - \frac{2}{\beta^{-}_\mu} ~\frac{1}{2 \pi \ii} \oint_C \dd{s} \frac{s^{|m|}}{(1-su_{-}) (1-su_{+})} \label{eq:app-int-is-s0} \\
&= - \frac{2}{u_{-} u_{+} \beta^{-}_\mu} ~\frac{1}{2 \pi \ii} \oint_C \dd{s} \frac{s^{|m|}}{(s - u^{-1}_{-}) (s - u^{-1}_{+})} \, , \label{eq:app-int-is-s}
\end{align} 
where $C$ is the same contour as in Eq.~\eqref{eq:app-int-modi}. Note that the substitution of $s$ in obtaining Eq.~\eqref{eq:app-int-is-s0} induces a negative sign, but it also reverses the direction of the  contour, thus maintaining the counterclockwise contour. Now the integrand in Eq.~\eqref{eq:app-int-is-s} has two simple poles at points $s=u^{-1}_{\pm}$, of which only the pole at $s=u^{-1}_{+}$ lies inside the closed contour $C$. Using the residue theorem once again, we obtain
\begin{align}
I_{m<0} = -\frac{2}{\beta^{-}_\mu} \frac{u^{-|m|}_{+}}{(u_{-} - u_{+})} = \frac{ f^{|m|}_{\mu} u^{|m|}_{-}}{\sqrt{1-\beta^{+}_\mu\beta^{-}_\mu}} \, .\label{eq:app-int-mg0}
\end{align}
For general $m \in \mathbb{Z}$, combining Eqs.~\eqref{eq:app-int-mleq0} and~\eqref{eq:app-int-mg0} one may write that
\begin{align}
I = \frac{ f^{-\frac{m-|m|}{2}}_{\mu} u^{|m|}_{-} }{\sqrt{1-\beta^{+}_\mu\beta^{-}_\mu}} = \frac{ f^{-\frac{m-|m|}{2}}_{\mu} }{\sqrt{1-\beta^{+}_\mu\beta^{-}_\mu}} \left( \frac{1 -  \sqrt{1-\beta^{+}_\mu\beta^{-}_\mu}}{\beta^{-}_\mu} \, \right)^{|m|} \, , \label{eq:app-int-res} 
\end{align}
which for $m=(n_0-n)$ yields Eq.~\eqref{eq:Inversion_integral2} in the main text.

\section{Solving the \tB Master equation}
\label{app:tB-solutions}

For an initial site $n_0 \in$ medium-1, i.e., for $n_0 \leq (M-1)$, the \tB Master equation \eqref{eq:master_eq_1_V}--\eqref{eq:master_eq_5_V} in the $z$-domain are given by
\begin{align}
& S_{1,1} (n,z|n_0) \left[ 1 -  z(1-q_1) \right] = \delta_{n,n_0} + \frac{z q_1}{2} (1-g_1) S_{1,1}(n-1,z|n_0) \nonumber \\
&\hskip4.5cm + \frac{z q_1}{2} (1+g_1) S_{1,1}(n+1,z|n_0)   \, ; \quad n < (M-1)  \, , \label{eq:master_eq_inZ_1_V} \\
& S_{1,1} (M-1,z|n_0) \left[ 1 -  z(1-q_1) \right] = \delta_{M-1,n_0} + \frac{z q_1}{2} (1-g_1) S_{1,1}(M-2,z|n_0) \nonumber \\
&\hskip5.3cm + \frac{z q_1}{2} (1+g_1) S_{c,1}(M,z|n_0) \, , \label{eq:master_eq_inZ_2_V} \\
& S_{c,1} (M,z|n_0) \!\left[ 1 - z \Big\{ 1- \frac{q_1}{2} (1+g_1) - \frac{q_2}{2} (1-g_2) \Big\} \right] =  \frac{z q_1 }{2} (1-g_1) S_{1,1}(M-1,z|n_0) \nonumber \\
&\hskip5.3cm + \frac{z q_2 }{2} (1+g_2) S_{2,1}(M+1,z|n_0)    \, , \label{eq:master_eq_inZ_3_V} \\
& S_{2,1} (M+1,z|n_0) \left[ 1 -  z(1-q_2) \right] =  \frac{z q_2}{2} (1-g_2) S_{c,1}(M,z|n_0) \nonumber \\
&\hskip5.3cm + \frac{z q_2}{2} (1+g_2) S_{2,1}(M+2,z|n_0) \, , \label{eq:master_eq_inZ_4_V} \\
& S_{2,1} (n,z|n_0) \left[ 1 -  z(1-q_2) \right] =  \frac{z q_2}{2} (1-g_2) S_{2,1}(n-1,z|n_0) \nonumber \\
&\hskip4.5cm + \frac{z q_2}{2} (1+g_2) S_{2,1}(n+1,z|n_0)  \, ; \quad n > (M+1) \, . \label{eq:master_eq_inZ_5_V} 
\end{align}
In this case, Eq.~\eqref{eq:Fourier_def_in_u} reduces to
\begin{align}
G_{1,1} (u,z|n_0) =\! \sum_{n=-\infty}^{M-1} u^n \, S_{1,1} (n,z|n_0)  ,  ~~
G_{2,1} (u,z|n_0) =\! \sum_{n=M+1}^{\infty} u^n \, S_{2,1} (n,z|n_0)  , \label{eq:GF_inU_2_def_V}
\end{align}
substituting Eqs.~\eqref{eq:master_eq_inZ_1_V}--\eqref{eq:master_eq_inZ_5_V} in which one obtains
\begin{align}
G_{1,1} (u,z|n_0) \left[ 1 - \frac{\beta^{-}_1}{2} u - \frac{\beta^{+}_1}{2}    u^{-1} \right] &= \frac{u^{n_0}}{1 -z(1-q_1)} - \frac{\beta^{-}_1 }{2} \, S_{1,1}(M-1,z|n_0) \, u^{M} \nonumber \\
&+  \frac{\beta^{+}_1 }{2} \,  S_{c,1} (M,z|n_0) \, u^{M-1} \, ,  \label{eq:GF_inU_1_V} \\
G_{2,1} (u,z|n_0) \left[ 1 - \frac{\beta^{-}_2}{2} u - \frac{\beta^{+}_2}{2}    u^{-1} \right] &= \frac{\beta^{-}_2   }{2 } \,  S_{c,1}(M,z|n_0) \, u^{M+1} \nonumber \\
&- \frac{\beta^{+}_2 }{2} \, S_{2,1}(M+1,z|n_0) \, u^{M} \, . \label{eq:GF_inU_2_V}
\end{align}
Using the inversion formula~\eqref{eq:Inverse_Fourier_def_in_u} and the integral relation~\eqref{eq:Inversion_integral2}, we obtain from Eqs.~\eqref{eq:GF_inU_1_V} and~\eqref{eq:GF_inU_2_V} that
\begin{align}
S_{1,1} (n,z|n_0) &= \frac{1}{\sqrt{1-\beta^{+}_1  \beta^{-}_1}} \Bigg[ \frac{ f_1^{ \frac{ n - n_0 + |n-n_0|}{2} } \xi_1^{|n-n_0|} }{1-z(1-q_1)} - \frac{\beta_1^{-}}{2}\, \xi_1^{M-n} S_{1,1}(M-1,z|n_0) \nonumber \\
&+ \frac{\beta_1^{+}}{2} \, \xi_1^{M-1-n} S_{c,1}(M,z|n_0)    \Bigg] \, , \label{eq:GF_inZ_1_V} 
\end{align}
and
\begin{align}
S_{2,1} (n,z|n_0) &= \frac{1}{\sqrt{1-\beta^{+}_2  \beta^{-}_2}} \Bigg[ \frac{\beta_2^{-}}{2} \, (f_2 \xi_2)^{n-M-1} S_{c,1}(M,z|n_0)  \nonumber \\
&- \frac{\beta_2^{+}}{2} \, (f_2 \xi_2)^{n-M} S_{2,1}(M+1,z|n_0)  \Bigg] \, , \label{eq:GF_inZ_2_V}
\end{align}
respectively.

By putting $n=(M-1)$ in Eq.~\eqref{eq:GF_inZ_1_V} and $n=(M+1)$ in Eq.~\eqref{eq:GF_inZ_2_V}, one gets
\begin{align}
& S_{1,1}(M-1, z|n_0) = \frac{ 2 f_1^{M-n_0 } \, \xi_1^{M-n_0}   }{ z q_1 (1-g_1)} + \xi_1 \, S_{c,1}(M,z|n_0)   \, , \label{eq:GF_inZ_S1S2_temp_V}
\end{align}
and
\begin{align}
& S_{2,1}(M+1, z|n_0) =  f_2 \, \xi_2 \, S_{c,1}(M,z|n_0)   \, , \label{eq:GF_inZ_S1S1_temp_V}
\end{align}
respectively. Equations~\eqref{eq:GF_inZ_S1S2_temp_V} and~\eqref{eq:GF_inZ_S1S1_temp_V} along with Eq.~\eqref{eq:master_eq_inZ_3_V} yield the solution $S_{c,1}(M,z|n_0)$ given in Eq.~\eqref{eq:GF_inZ_Sc_V}.
Substituting Eq.~\eqref{eq:GF_inZ_Sc_V} back in Eqs.~\eqref{eq:GF_inZ_S1S2_temp_V} and~\eqref{eq:GF_inZ_S1S1_temp_V}, we obtain
\begin{align}
S_{1,1}(M-1, z|n_0) &= \frac{ 2 f_1^{M-n_0 } \, \xi_1^{M-n_0}   }{ z q_1 (1-g_1)} + \frac{1}{\Gamma} \, f_1^{M-n_0} \xi_1^{M-n_0+1}   \, ,  \label{eq:GF_inZ_S1_temp_sol_V} 
\end{align}
and
\begin{align}
S_{2,1}(M+1, z|n_0) &=   \frac{1}{\Gamma} \, f_1^{M-n_0} \xi_1^{M-n_0} \,f_2 \, \xi_2 \, , \label{eq:GF_inZ_S2_temp_sol_V}
\end{align}
respectively. Finally, substituting Eqs.~\eqref{eq:GF_inZ_Sc_V},~\eqref{eq:GF_inZ_S1_temp_sol_V}, and~\eqref{eq:GF_inZ_S2_temp_sol_V} in Eqs.~\eqref{eq:GF_inZ_1_V}--\eqref{eq:GF_inZ_2_V}, one obtains Eqs.~\eqref{eq:GF_inZ_S1_sol_V} and ~\eqref{eq:GF_inZ_S2_sol_V} in the main text.

The \tB generating function $S_{\mu,\nu}(n, z|n_0)$ for initial conditions $n_0=M$ and $n_0\geq (M+1)$ may be obtained by following the same procedure.
When the walker starts exactly from the interface, i.e., for $n_0 = M$, we get
\begin{align}
S_{1,c}(n, z|n_0) = \frac{1}{\Gamma} \, \xi_1^{M-n} ,  ~\,
S_{c,c}(M,z|n_0) = \frac{1}{\Gamma}   , ~\,
S_{2,c}(n, z|n_0) = \frac{1}{\Gamma} \, f_2^{n-M} \, \xi_2^{n-M}   .  \label{eq:GF_inZ_S1c2_temp_sol_V_regC}
\end{align}
On the other hand, for $n_0 \geq (M+1)$, one obtains
\begin{align}
& S_{1,2}(n, z|n_0) = \frac{1}{\Gamma} \, \xi_1^{M-n} \, \xi_2^{n_0-M} \,  ,  \label{eq:GF_inZ_S1_temp_sol_V_regIII} \\
& S_{c,2}(M,z|n_0)  = \frac{1}{\Gamma} \, \xi_2^{n_0-M} \, ,  \label{eq:GF_inZ_Sc_V_regIII} \\
& S_{2,2}(n, z|n_0) = \frac{1}{\sqrt{1-\beta^{+}_2  \beta^{-}_2}} \Bigg[ \frac{ f_2^{ \frac{ n - n_0 + |n-n_0|}{2} } \xi_2^{|n-n_0|} }{1-z+ z q_2} \nonumber \\
& \hskip2.6cm - f_2^{n-M} \, \xi_2^{n+n_0-2M} \bigg\{ \frac{1}{1-z+ z q_2} - \frac{1}{\Gamma} \, \sqrt{1-\beta^{+}_2  \beta^{-}_2} \bigg\} \Bigg]  \, .  \label{eq:GF_inZ_S2_temp_sol_V_regIII}
\end{align}

\section{Steady-state without bias in reflecting domain}
\label{app:bounded-steady-state}

We rewrite the left-bounded generating function from Eq.~\eqref{eq:Defect-1-LB-inhomo} as
\begin{align}
L^{(\mathrm{r})}_{\mu,\nu }(n,z|n_0) = S_{\mu,\nu }(n,z|n_0) + \frac{\mathbb{N}_{\mu}(n,z)}{\mathbb{D}(z)} S_{1,\nu }(1,z|n_0) \, ; \quad n\geq 1 \, ,
\label{eq:Lr}
\end{align}
where
\begin{align}
& \mathbb{N}_{\mu}(n,z) \equiv S_{\mu,1 }(n,z|1) - S_{\mu,1 }(n,z|0)  \, ; \quad n\geq 1 \, , \label{eq:Nmu} \\
& \mathbb{D}(z) \equiv \frac{2}{z q_1 (1+g_1)} - S_{1,1 }(1,z|1)   +   S_{1,1 }(1,z|0) \, . \label{eq:D}
\end{align}
In the steady state, for both \tA and \tB dynamics without bias, the semi-bounded propagator $L^{(\mathrm{r})}_{\mu,\nu }(n,z|n_0)$ vanishes, as may be seen by computing the steady-state probability 
\begin{align}
{}^{\mathrm{ss}}L^{(\mathrm{r})}_{\mu,\nu }(n|n_0) &\equiv \lim_{z\to 1} (1-z) L^{(\mathrm{r})}_{\mu,\nu }(n,z|n_0) \nonumber \\
&= \lim_{z\to 1} (1-z) \Big[ S_{\mu,\nu }(n,z|n_0) + \frac{\mathbb{N}_{\mu}(n,z)}{\mathbb{D}(z)} S_{1,\nu }(1,z|n_0) \Big] \, . \label{eq:Lss}
\end{align}
Using Eqs.~\eqref{eq:GF_inZ_1_sol}--\eqref{eq:GF_inZ_2_sol_Reg2} with $g_\mu=0$, and expanding the functions $S_{\mu,\nu }(n,z|n_0)$, $\mathbb{N}_{\mu}(n,z)$, and $\mathbb{D}(z)$ in powers of $(1-z)$, one obtains for \tA dynamics that 
\begin{align}
& S_{\mu,\nu }(n,z|n_0) = \frac{\sqrt{2 q_1 q_2}}{q_\mu \big(\sqrt{q_1} + \sqrt{q_2}\big)  \sqrt{1-z}} - \frac{|n-n_0|}{q_\mu} + \frac{(1+2M-n-n_0)\big(\sqrt{q_1}-\sqrt{q_2}\big)}{q_\mu \big(\sqrt{q_1} + \sqrt{q_2}\big)} \nonumber \\
&\hskip70pt + \mathcal{O}(\sqrt{1-z}) \, ;\quad \mu,\nu \in \{1,2\} \, , \label{eq:Smunu-tA-lim}\\
& \mathbb{N}_{\mu}(n,z) = \frac{  \sqrt{4 q_2}}{q_\mu \big(\sqrt{q_1} + \sqrt{ q_2} \big) }+ \mathcal{O}(\sqrt{1-z})  \, ;\quad \mu \in \{1,2\} \, , \label{eq:Nmu-tA-lim} \\
& \mathbb{D}(z) = \frac{2}{\sqrt{q_1} (\sqrt{q_1} + \sqrt{q_2})} + \frac{\sqrt{2} \big\{ (1-2M)\sqrt{q_1}+2M \sqrt{q_2} \big\} }{q_1 \sqrt{q_1} (\sqrt{q_1} + \sqrt{q_2})} \sqrt{1-z} + \mathcal{O}({1-z})  \, . \label{eq:D-tA-lim}
\end{align}
Similarly, using Eqs.~\eqref{eq:GF_inZ_S1_sol_V}--\eqref{eq:GF_inZ_S2_sol_V} and~\eqref{eq:GF_inZ_S1c2_temp_sol_V_regC}--\eqref{eq:GF_inZ_S2_temp_sol_V_regIII}, for \tB dynamics one obtains
\begin{align}
& S_{\mu,\nu }(n,z|n_0) = \frac{\sqrt{2}}{\big(\sqrt{q_1} + \sqrt{q_2} \big) \sqrt{1-z}} + h_{\mu,\nu}(n,n_0)+ \mathcal{O}(\sqrt{1-z})  \, ,  \label{eq:Smunu-tB-lim} \\
& \mathbb{N}_{\mu}(n,z) = \frac{ 2}{\sqrt{q_1} (\sqrt{q_1} + \sqrt{q_2})} + \mathcal{O}(\sqrt{1-z})  \, ;\quad \mu \in \{1,c,2\}\, ,  \label{eq:Nmu-tB-lim} \\
& \mathbb{D}(z) = \frac{ \sqrt{4 q_2}}{q_1(\sqrt{q_1}+\sqrt{ q_2})}+ \mathcal{O}(\sqrt{1-z})  \, , \label{eq:D-tB-lim}
\end{align}
where
\begin{align}
h_{\mu,\nu}(n,n_0) \equiv 
\begin{cases} 
- \dfrac{|n-n_0|}{q_\mu} - \left( \dfrac{\sqrt{q_1}-\sqrt{q_2}}{\sqrt{q_1}+\sqrt{q_2}}\right) \dfrac{2 M-n - n_0}{q_\mu} &  \quad \mu=\nu; ~\mu \in \{1,c,2\} \, , \\[3ex]
- \dfrac{2|M-n|}{q_\mu + \sqrt{q_1 q_2}}  -\dfrac{2|M-n_0|}{q_\nu + \sqrt{q_1 q_2}} &  \quad \mu \neq \nu; ~\mu,\nu \in \{1,c,2\} \, .
   \end{cases} \label{eq:hmunu}
\end{align}
Substituting Eqs.~\eqref{eq:Smunu-tA-lim}--\eqref{eq:D-tB-lim} in Eq.~\eqref{eq:Lss}, one may easily check that ${}^{\mathrm{ss}}L^{(\mathrm{r})}_{\mu,\nu }(n|n_0) = 0$ for both \tA and \tB dynamics.

The fully-bounded propagator in Eq.~\eqref{eq:Defect-1-FB-inhomo} may be rewritten as
\begin{align}
F^{\mathrm{(r)}}_{\mu, \nu}(n,z|n_0) &=  L^{\mathrm{(r)}}_{\mu, \nu}(n,z|n_0)   + \frac{\mathbb{M}_{\mu}(n,z) \, \mathbb{R}_{\nu}(n_0,z)}{\mathbb{C}(z)} \, ,
\end{align}
where
\begin{align}
& \mathbb{M}_{\mu}(n,z) \equiv L^{(\mathrm{r})}_{\mu,2 }(n,z|N) - L^{(\mathrm{r})}_{\mu,2 }(n,z|N+1)  =  S_{\mu,2 }(n,z|N)  - S_{\mu,2 }(n,z|N+1) \nonumber \\
&\hskip40pt + \frac{\mathbb{N}_{\mu}(n,z)}{\mathbb{D}(z)} \big[ S_{1,2 }(1,z|N)  -  S_{1,2 }(1,z|N+1) \big]    \, , \label{eq:def-M-mu-n-z} \\
& \mathbb{C}(z) \equiv \frac{2}{z q_2 (1-g_2)} - \mathbb{M}_{2}(N,z) \, ,  \label{eq:def-C-z}  \\
& \mathbb{R}_{\nu}(n_0,z) \equiv   L^{(\mathrm{r})}_{2,\nu }(N,z|n_0)  \, . \label{eq:def-R-nu-n0-z}
\end{align}
In the fully-bounded domain, the steady-state probability is then given by 
\begin{align}
{}^{\mathrm{ss}}F^{\mathrm{(r)}}_{\mu, \nu}(n|n_0) \equiv  \lim_{z\to 1} (1-z) F^{\mathrm{(r)}}_{\mu, \nu}(n,z|n_0) = \lim_{z\to 1} \frac{(1-z) \, \mathbb{M}_{\mu}(n,z) \, \mathbb{R}_{\nu}(n_0,z)}{\mathbb{C}(z)}  \, ,  \label{eq:FB-SS-1}
\end{align}
where we have used ${}^{\mathrm{ss}}L^{(\mathrm{r})}_{\mu,\nu }(n|n_0) = 0$. To compute the limit in Eq.~\eqref{eq:FB-SS-1}, one may expand the functions $\mathbb{M}_{\mu}(n,z)$,      $\mathbb{R}_{\nu}(n_0,z)$, and ${\mathbb{C}(z)}$ in powers of $(1-z)$.
For \tA dynamics without bias, using Eqs.~\eqref{eq:GF_inZ_1_sol}--\eqref{eq:GF_inZ_2_sol_Reg2} one obtains
\begin{align}
& \mathbb{M}_{\mu}(n,z) = \frac{2}{q_\mu} +  \mathcal{O}(\sqrt{1-z})   \, ;\quad \mu \in \{1,2\} \,, \label{eq:FB-SS-M1} \\
& \mathbb{R}_{\nu}(n_0,z) = {\frac{\sqrt{2}}{\sqrt{q_2 (1-z)}}} - \frac{(2N-2M-1)q_1+2Mq_2}{q_1 q_2} + \mathcal{O}(\sqrt{1-z})  \,;\quad \nu \in \{1,2\} \, , \label{eq:FB-SS-R1}\\
& \mathbb{C}(z) = \frac{2\sqrt{2}}{\sqrt{q_2}} \left( \frac{M}{q_1} + \frac{ N-M}{q_2} \right)  \sqrt{1-z} + \mathcal{O}(1-z)\, .\label{eq:FB-SS-C1}
\end{align}
Similarly, for \tB dynamics without bias, using Eqs.~\eqref{eq:GF_inZ_S1_sol_V}--\eqref{eq:GF_inZ_S2_sol_V} and~\eqref{eq:GF_inZ_S1c2_temp_sol_V_regC}--\eqref{eq:GF_inZ_S2_temp_sol_V_regIII} one obtains
\begin{align}
& \mathbb{M}_{\mu}(n,z) = \frac{2}{q_2} +  \mathcal{O}(\sqrt{1-z})  \, ;\quad \mu \in \{1,c,2\} \, , \label{eq:FB-SS-M1-TB} \\
& \mathbb{R}_{\nu}(n_0,z) = {\frac{\sqrt{2}}{\sqrt{q_2 (1-z)}}} - \frac{2N-1}{q_2} + \mathcal{O}(\sqrt{1-z})  \, ;\quad \nu \in \{1,c,2\} \, , \label{eq:FB-SS-R1-TB}\\
& \mathbb{C}(z) = \frac{2\sqrt{2} \, N}{q_2\sqrt{q_2}}  \sqrt{1-z} + \mathcal{O}(1-z)\, .\label{eq:FB-SS-C1-TB}
\end{align}
Equations~\eqref{eq:FB-SS-M1}--\eqref{eq:FB-SS-C1-TB} when substituted in Eq.~\eqref{eq:FB-SS-1} yield Eq.~\eqref{eq:ssTAB} of the main text.

\section{Mean first-passage time without bias in reflecting domain}
\label{app:bounded-FP}

The mean first-passage time from site $n_0$ in medium-$\nu$ to site $n$ in medium-$\mu$ obtained by using the relation $T^{\mathrm{(r)}}_{\substack{{n_0 \to n} \\ {\nu \to \mu}}} \equiv \lim_{z \to 1} \pdv{z}\mathcal{F}^{\mathrm{(r)}}_{\mu,\nu}(n,z|n_0)$ is given by
\begin{align}
T^{\mathrm{(r)}}_{\substack{{n_0 \to n} \\ {\nu \to \mu}}} = \lim_{z \to 1} ~\frac{\mathfrak{N}_{\mu,\nu}(z,n,n_0)}{\mathfrak{D}_{\mu}(z,n)} \, , \label{eq:MFPT-def}
\end{align}
where we have
\begin{align}
& \mathfrak{N}_{\mu,\nu}(z,n,n_0) \equiv  \big\{ \mathbb{C}(z) L^{\mathrm{(r)}}_{\mu, \mu}(n,z|n)   + \mathbb{M}_{\mu}(n,z) \, \mathbb{R}_{\mu}(n,z) \big\} \big\{ \mathbb{C}'(z) L^{\mathrm{(r)}}_{\mu, \nu}(n,z|n_0) \nonumber \\
&\hskip60pt  +\mathbb{C}(z) L^{\mathrm{(r)}'}_{\mu, \nu}(n,z|n_0)   + \mathbb{M}'_{\mu}(n,z) \, \mathbb{R}_{\nu}(n_0,z)+  \mathbb{M}_{\mu}(n,z) \, \mathbb{R}'_{\nu}(n_0,z)) \big\} \nonumber \\
&\hskip60pt -   \big\{ \mathbb{C}(z) L^{\mathrm{(r)}}_{\mu, \nu}(n,z|n_0)   + \mathbb{M}_{\mu}(n,z) \, \mathbb{R}_{\nu}(n_0,z) \big\}
 \big\{  \mathbb{C}'(z) L^{\mathrm{(r)}}_{\mu, \mu}(n,z|n) \nonumber \\
&\hskip60pt + \mathbb{C}(z) L^{\mathrm{(r)}'}_{\mu, \mu}(n,z|n)  + \mathbb{M}'_{\mu}(n,z) \, \mathbb{R}_{\mu}(n,z)  + \mathbb{M}_{\mu}(n,z) \, \mathbb{R}'_{\mu}(n,z)  \big\} \, , \label{eq:dFdz-numer} \\
& \mathfrak{D}_{\mu}(z,n) \equiv  \big[ \mathbb{C}(z) L^{\mathrm{(r)}}_{\mu, \mu}(n,z|n)   + \mathbb{M}_{\mu}(n,z) \, \mathbb{R}_{\mu}(n,z) \big]^2 \, . \label{eq:dFdz-denom}
\end{align}

To compute the limit in Eq.~\eqref{eq:MFPT-def}, we again expand the relevant functions in a power series of $(1-z)$, and the results may generally be written as 
\begin{align}
& L^{(\mathrm{r})}_{\mu,\nu }(n,z|n_0) \equiv l^{(1)}_{\mu,\nu}(n,n_0) (1-z)^{-1/2} + l^{(2)}_{\mu,\nu}(n,n_0) (1-z)^{0} \nonumber \\
& \hspace{2cm} + l^{(3)}_{\mu,\nu}(n,n_0) (1-z)^{1/2}+\ldots  \, , \label{eq:Llead} \\
& \mathbb{M}_\mu(n,z) \equiv m^{(1)}_\mu(n) (1-z)^0 + m^{(2)}_\mu(n) (1-z)^{1/2} + m^{(3)}_\mu(n) (1-z)+\ldots \, , \label{eq:Mlead}\\
& \mathbb{C}(z) \equiv c^{(1)} (1-z)^{1/2} + c^{(2)} (1-z)^{1} + c^{(3)} (1-z)^{3/2}+\ldots \, , \label{eq:Clead}\\
& \mathbb{R}_{\nu }(n_0,z) \equiv r^{(1)}_{\nu}(n_0) (1-z)^{-1/2} + r^{(2)}_{\nu}(n_0) (1-z)^{0} + r^{(3)}_{\nu}(n_0) (1-z)^{1/2}+\ldots   . \label{eq:Rlead}
\end{align}
Similar expansion of derivatives of the above functions with respect to $z$ yields
\begin{align}
& L^{(\mathrm{r})'}_{\mu,\nu }(n,z|n_0) \equiv {\overline{l}}^{(1)}_{\mu,\nu}(n,n_0) (1-z)^{-3/2} + {\overline{l}}^{(2)}_{\mu,\nu}(n,n_0) (1-z)^{-1} \nonumber \\
& \hspace{2cm} + {\overline{l}}^{(3)}_{\mu,\nu}(n,n_0) (1-z)^{-1/2}+\ldots \, , \label{eq:DLlead}\\
& \mathbb{M}'_\mu(n,z) \equiv {\overline{m}}^{(1)}_\mu(n) (1-z)^{-1/2} + {\overline{m}}^{(2)}_\mu(n) (1-z)^{0} + {\overline{m}}^{(3)}_\mu(n) (1-z)^{1/2}+\ldots \, , \label{eq:DMlead}\\
& \mathbb{C}'(z) \equiv {\overline{c}}^{(1)} (1-z)^{-1/2} + {\overline{c}}^{(2)} (1-z)^{0} + {\overline{c}}^{(3)} (1-z)^{1/2}+\ldots \, , \label{eq:DClead} \\
& \mathbb{R}'_{\nu }(n_0,z) \equiv {\overline{r}}^{(1)}_{\nu}(n_0) (1-z)^{-3/2} + {\overline{r}}^{(2)}_{\nu}(n_0) (1-z)^{-1} + {\overline{r}}^{(3)}_{\nu}(n_0) (1-z)^{-1/2}+\ldots   . \label{eq:DRlead}
\end{align}
The leading-order behavior in $(1-z)$ of the functions in Eqs.~\eqref{eq:Llead}--\eqref{eq:DRlead} are the same for both \tA and \tB dynamics, while the coefficients of the terms of each order may differ for the two dynamics. For example, we have $m^{(1)}_\mu(n) = 2/q_\mu$ and $2/q_2$ for \tA and \tB dynamics, respectively, while $r^{(1)}_\nu(n_0) = \sqrt{2/q_2}$ for both the dynamics (see Eqs.~\eqref{eq:FB-SS-M1}--\eqref{eq:FB-SS-R1} and~\eqref{eq:FB-SS-M1-TB}--\eqref{eq:FB-SS-R1-TB}).

As $z \to 1$, using Eqs.~\eqref{eq:Llead}--\eqref{eq:DRlead}, we obtain to the leading-order from Eq.~\eqref{eq:dFdz-denom}  that
\begin{align}
\mathfrak{D}_{\mu}(z,n) = \frac{\big[  m^{(1)}_\mu(n) ~  r^{(1)}_\mu(n) \big]^2}{1-z} \, ,\label{eq:dFdz-denom1}
\end{align} 
and similarly from Eq.~\eqref{eq:dFdz-numer} upto the third order that
\begin{align}
\mathfrak{N}_{\mu,\nu}(z,n,n_0) = \frac{A_{\mu,\nu}(n,n_0)}{(1-z)^2} + \frac{B_{\mu,\nu}(n,n_0)}{(1-z)^{3/2}} + \frac{C_{\mu,\nu}(n,n_0)}{1-z} \, ,\label{eq:dFdz-numer1}
\end{align}
where
\begin{align}
A_{\mu,\nu}(n,n_0) &\equiv  \big( m^{(1)}_\mu(n) \big)^2 \, \,\big[  r^{(1)}_\mu(n) \, {\overline{r}}^{(1)}_\nu(n_0) -  {\overline{r}}^{(1)}_\mu(n)  \, r^{(1)}_\nu(n_0)  \big] \, , \\
B_{\mu,\nu}(n,n_0) &\equiv m^{(1)}_\mu(n) \Big[  
  r^{(1)}_\mu(n)   \big\{ c^{(1)} {\overline{l}}^{(1)}_{\mu,\nu}(n,n_0) + {\overline{c}}^{(1)} l^{(1)}_{\mu,\nu}(n,n_0) + m^{(1)}_\mu(n) {\overline{r}}^{(2)}_{\nu}(n_0)  \big\}  \nonumber \\
& - r^{(1)}_\nu(n_0) \big\{ c^{(1)} {\overline{l}}^{(1)}_{\mu,\mu}(n,n)   + {\overline{c}}^{(1)} l^{(1)}_{\mu,\mu}(n,n) + m^{(1)}_\mu(n) {\overline{r}}^{(2)}_{\mu}(n)    \big\}   \nonumber \\
& - {\overline{r}}^{(1)}_\mu(n)   \big\{ c^{(1)} l^{(1)}_{\mu,\nu}(n,n_0) + 2 m^{(2)}_\mu(n) r^{(1)}_\nu(n_0)  + m^{(1)}_\mu(n) r^{(2)}_\nu(n_0)\big\} \nonumber \\
& + {\overline{r}}^{(1)}_\nu(n_0) \big\{ c^{(1)} l^{(1)}_{\mu,\mu}(n,n)   + 2 m^{(2)}_\mu(n) r^{(1)}_\mu(n)    + m^{(1)}_\mu(n) r^{(2)}_\mu(n)  \big\} \Big] \, ,
\end{align}
and 
\begin{align}
C_{\mu,\nu}(n,n_0) &\equiv m^{(1)}_\mu(n) \Big[ 
  r^{(1)}_\mu(n) \Big\{ c^{(2)} {\overline{l}}^{(1)}_{\mu,\nu}(n,n_0) + c^{(1)} {\overline{l}}^{(2)}_{\mu,\nu}(n,n_0) + 
    {\overline{c}}^{(2)} l^{(1)}_{\mu,\nu}(n,n_0)  \nonumber \\
&+ {\overline{c}}^{(1)} l^{(2)}_{\mu,\nu}(n,n_0)   +     {\overline{r}}^{(3)}_{\nu}(n_0) m^{(1)}_\mu(n) + {\overline{r}}^{(2)}_{\nu}(n_0) m^{(2)}_\mu(n) + 
    {\overline{r}}^{(1)}_{\nu}(n_0) m^{(3)}_\mu(n) \nonumber \\
&+ {\overline{m}}^{(2)}_\mu(n) r^{(1)}_\nu(n_0) + 
    {\overline{m}}^{(1)}_\mu(n) r^{(2)}_\nu(n_0) \Big\} -r^{(1)}_\nu(n_0) \Big\{ c^{(2)} {\overline{l}}^{(1)}_{\mu,\mu}(n,n)  \nonumber \\ 
&+ c^{(1)} {\overline{l}}^{(2)}_{\mu,\mu}(n,n) + 
    {\overline{c}}^{(2)} l^{(2)}_{\mu,\mu}(n,n) + {\overline{c}}^{(1)} l^{(2)}_{\mu,\mu}(n,n) +     {\overline{r}}^{(3)}_{\mu}(n) m^{(1)}_\mu(n)  \nonumber \\
& + {\overline{r}}^{(2)}_{\mu}(n) m^{(2)}_\mu(n) + 
    {\overline{r}}^{(1)}_{\mu}(n) m^{(3)}_\mu(n) + {\overline{m}}^{(2)}_\mu(n) r^{(1)}_\mu(n) + 
    {\overline{m}}^{(1)}_\mu(n) r^{(2)}_\mu(n) \Big\} \nonumber \\ 
&+{\overline{r}}^{(1)}_{\nu}(n_0) \Big\{ c^{(2)} l^{(1)}_{\mu,\mu}(n,n) + c^{(1)} l^{(2)}_{\mu,\mu}(n,n) + 
    m^{(3)}_\mu(n) r^{(1)}_\mu(n) \nonumber \\
& + m^{(2)}_\mu(n) r^{(2)}_\mu(n) + 
    m^{(1)}_\mu(n) r^{(3)}_\mu(n) \Big\} -{\overline{r}}^{(1)}_{\mu}(n) \Big\{ c^{(2)} l^{(1)}_{\mu,\nu}(n,n_0)  \nonumber \\
& + c^{(1)} l^{(2)}_{\mu,\nu}(n,n_0) + 
    m^{(3)}_\mu(n) r^{(1)}_\nu(n_0) + m^{(2)}_\mu(n) r^{(2)}_\nu(n_0) + 
    m^{(1)}_\mu(n) r^{(3)}_\nu(n_0) \Big\}
\Big] \nonumber \\
&+ \Big\{ c^{(1)} l^{(1)}_{\mu,\mu}(n,n)   + m^{(2)}_\mu(n) r^{(1)}_\mu(n) + m^{(1)}_\mu(n) r^{(2)}_\mu(n) \Big\} \times \Big\{ c^{(1)} {\overline{l}}^{(1)}_{\mu,\nu}(n,n_0)   \nonumber \\
&+  {\overline{c}}^{(1)} l^{(1)}_{\mu,\nu}(n,n_0) + {\overline{r}}^{(2)}_{\nu}(n_0) m^{(1)}_\mu(n) + 
    {\overline{r}}^{(1)}_{\nu}(n_0) m^{(2)}_\mu(n) + {\overline{m}}^{(1)}_\mu(n) r^{(1)}_\nu(n_0) \Big\}    \nonumber \\
&- \Big\{ c^{(1)} l^{(1)}_{\mu,\nu}(n,n_0)  + m^{(2)}_\mu(n) r^{(1)}_\nu(n_0) + m^{(1)}_\mu(n) r^{(2)}_\nu(n_0) \Big\} \times \Big\{ c^{(1)} {\overline{l}}^{(1)}_{\mu,\mu}(n,n)    \nonumber \\
&+  {\overline{c}}^{(1)} l^{(1)}_{\mu,\mu}(n,n)  + {\overline{r}}^{(2)}_{\mu}(n) m^{(1)}_\mu(n) + 
    {\overline{r}}^{(1)}_{\mu}(n) m^{(2)}_\mu(n) + {\overline{m}}^{(1)}_\mu(n) r^{(1)}_\mu(n) \Big\}    \, .
\end{align}

For both \tA and \tB dynamics, the first-order coefficients ${l}^{(1)}_{\mu,\nu}(n,n_0)$, ${\overline{l}}^{(1)}_{\mu,\nu}(n,n_0)$ for all possible values of $\mu,\nu$ are independent of the arguments $n,n_0$ and the index $\nu$, i.e., we have ${l}^{(1)}_{\mu,\nu}(n,n_0)={l}^{(1)}_{\mu}$, ${\overline{l}}^{(1)}_{\mu,\nu}(n,n_0)={\overline{l}}^{(1)}_{\mu}$. As a result, one has $r^{(1)}_\mu(n) = r^{(1)}_\nu(n_0) = r^{(1)}$, $ {\overline{r}}^{(1)}_\mu(n) = {\overline{r}}^{(1)}_\nu(n_0) = {\overline{r}}^{(1)}$, which yields $A_{\mu,\nu}(n,n_0)=0$ for both dynamics. On the other hand, while the second-order coefficient ${l}^{(2)}_{\mu,\nu}(n,n_0)$ does depend on its arguments, the coefficient $r^{(2)}_\mu(n)$ is independent of $n$ and $\mu$ for both dynamics, i.e., $r^{(2)}_\mu(n)=r^{(2)}_\nu(n_0)=r^{(2)}$. Additionally, for both the dynamics we get ${\overline{l}}^{(2)}_{\mu,\nu}(n,n_0)=0$ and ${\overline{r}}^{(2)}_{\mu}(n)=0$. These relations not only yield $B_{\mu,\nu}(n,n_0)=0$, but also greatly simplify $C_{\mu,\nu}(n,n_0)$, and finally, from Eqs.~\eqref{eq:MFPT-def},~\eqref{eq:dFdz-denom1}, and~\eqref{eq:dFdz-numer1} one obtains for both \tA and \tB dynamics that
\begin{align}
T^{\mathrm{(r)}}_{\nu \to \mu} &= \frac{ \big( c^{(1)} \, {\overline{r}}^{(1)} - {\overline{c}}^{(1)} \, r^{(1)} \big) \big[ l^{(2)}_{\mu,\mu}(n,n) - l^{(2)}_{\mu,\nu}(n,n_0) \big]}{   (r^{(1)})^2 ~ m^{(1)}_\mu }  +   \frac{ {\overline{r}}^{(1)} \big[ r^{(3)}_\mu(n)-r^{(3)}_\nu(n_0) \big] }{ (r^{(1)})^2 } \nonumber \\ 
&  - \frac{ {\overline{r}}^{(3)}_\mu(n)-{\overline{r}}^{(3)}_\nu(n_0)  }{ r^{(1)}} \, , \label{eq:MFPT-Final}
\end{align}
where we have also used the fact that $m^{(1)}_\mu(n)$ is independent $n$ (see Eqs.~\eqref{eq:FB-SS-M1} and~\eqref{eq:FB-SS-M1-TB}). Substituting the coefficients for all possible values of $\mu$ and $\nu$ on the right-hand side of Eq.~\eqref{eq:MFPT-Final} for both \tA and \tB, one finally obtains the MFPT given in Eqs.~\eqref{eq:MFPT-refl-TA1}--\eqref{eq:MFPT-refl-TB} of the main text.

\section{Propagator in the continuous limit}
\label{app:CSCT-Reg2}

In heterogeneous space with $x_0 \in $ medium-$2$, the CSCT propagators obtained using Eqs.~\eqref{eq:DSCT-gen},~\eqref{eq:GF_inZ_1_sol_Reg2},~\eqref{eq:GF_inZ_2_sol_Reg2},~\eqref{eq:GF_inZ_S1_temp_sol_V_regIII}, and~\eqref{eq:GF_inZ_S2_temp_sol_V_regIII} are given by 
\begin{align}
\label{eq:CSCT-12}
& \widetilde{P}_{1,2}(x,\epsilon|x_0) = \frac{2 \rho \exp\Big[{ \frac{(x_M-x)}{2 D_1}(\gamma_1 - \sqrt{\gamma^2_1 + 4 \epsilon D_1}) -\frac{(x_M-x_0)}{2 D_2}(\gamma_2 - \sqrt{\gamma^2_2 + 4 \epsilon D_2}) \Big]} }{ \rho \big(  \gamma_1 +\sqrt{\gamma^2_1 + 4\epsilon D_1} \big)  - \gamma_2 + \sqrt{\gamma^2_2 + 4\epsilon D_2} } \, , \\
\label{eq:CSCT-22}
& \widetilde{P}_{2,2}(x,\epsilon|x_0) = \frac{\ee^{-\frac{(x-x_0)\gamma_2}{2 D_2}}   }{\sqrt{\gamma^2_2 + 4 \epsilon D_2}} \Bigg[ \exp\Big\{{-\frac{|x-x_0|}{2 D_2} \sqrt{\gamma^2_2 + 4 \epsilon D_2}} \Big\} \nonumber \\[1ex]
& -  \frac{ \rho \big(\gamma_1 +\sqrt{\gamma^2_1 + 4\epsilon D_1} \, \big) -  \gamma_2 -\sqrt{\gamma^2_2 + 4\epsilon D_2}   }{ \rho \big(\gamma_1 +\sqrt{\gamma^2_1 + 4\epsilon D_1} \, \big) -  \gamma_2 +\sqrt{\gamma^2_2 + 4\epsilon D_2}  }  \exp\Big\{{- \frac{(x+x_0-2 x_M)}{2 D_2} \sqrt{\gamma^2_2 + 4 \epsilon D_2}} \Big\}  \Bigg]  ,
\end{align}
with $\rho$ defined in Eq.~\eqref{eq:rho-def}.

In bias-less ($\gamma_\mu=0$) heterogeneous space with $x_0 \in $ medium-$1$, the propagators obtained from Eqs.~\eqref{eq:CSCT-11}--\eqref{eq:CSCT-21} reduce to
\begin{align}
\label{eq:CSCT-11-g0}
\widetilde{P}_{1,1}(x,\epsilon|x_0) &= \frac{1}{\sqrt{4 \epsilon D_1}} \Bigg[ \ee^{-|x-x_0| \sqrt{\frac{ \epsilon }{ D_1}}}  -  \frac{  \sqrt{ D_2}   -\rho \sqrt{ D_1} }{ \sqrt{ D_2} + \rho \sqrt{ D_1}  }  \,\ee^{- (2 x_M -x -x_0)\sqrt{\frac{ \epsilon }{ D_1}}}  \Bigg] \, , \\
\label{eq:CSCT-21-g0}
\widetilde{P}_{2,1}(x,\epsilon|x_0) &= \frac{ 1}{ (\sqrt{ D_2} + \rho \sqrt{ D_1}) \sqrt{\epsilon}  }   \,\exp\!\left[ - \sqrt{\epsilon} \left( \dfrac{x_M-x_0}{\sqrt{D1}} + \dfrac{x-x_M}{\sqrt{D2}} \right) \right] \, ,
\end{align}
which when inverted yield the time-dependent propagators
\begin{align}
\label{eq:CSCT-11-g0-int}
{P}_{1,1}(x,\tau|x_0) &= \frac{1}{\sqrt{4 \pi D_1 \tau}} \Bigg[ \ee^{-\frac{(x-x_0)^2}{4D_1 \tau} }  -  \frac{  \sqrt{ D_2}   -\rho \sqrt{ D_1} }{ \sqrt{ D_2} + \rho \sqrt{ D_1}  }  \, \ee^{- \frac{(2 x_M -x -x_0)^2} {4 D_1 \tau}}  \Bigg] \, , \\
\label{eq:CSCT-21-g0-int}
{P}_{2,1}(x,\tau|x_0) &= \frac{ 1}{ (\sqrt{ D_2} + \rho \sqrt{ D_1}) \sqrt{\pi \tau}  }  \,\exp\left[ - \dfrac{1}{4 \tau} \left( \dfrac{x_M-x_0}{\sqrt{D1}} + \dfrac{x-x_M}{\sqrt{D2}} \right)^2 \right] \, .
\end{align}

\section*{References}
\bibliographystyle{unsrt} 
\bibliography{library}

\end{document}